\documentclass[twocolumn]{aastex631}
\usepackage{CJK}
\shorttitle{Her X-1}
\shortauthors{Chou \& Wu}
\begin{document}
\begin{CJK*}{Bg5}{bsmi}
\title{Superorbital Phase Evolution and Identification of New Anomalous Low State Candidates for Her X-1 }

\correspondingauthor{Yi Chou}
\email{yichou@astro.ncu.edu.tw}

\author[0000-0002-8584-2092]{Yi Chou (©PÖö)}
\affiliation{Graduate Institute of Astronomy, National Central University \\
  300 Jhongda Rd. Jhongli Dist. Taoyuan, 320317, Taiwan}
\author[0009-0006-9517-491X]{Jun-Lei Wu (§d§g½U)}
\affiliation{Graduate Institute of Astronomy, National Central University \\
  300 Jhongda Rd. Jhongli Dist. Taoyuan, 320317, Taiwan}



\begin{abstract}

The 35-day superorbital modulation of the X-ray binary Her X-1 was systematically investigated using a 35-year multi-instrument X-ray observational baseline. We developed a Rank-Searching method using the peak intensity of the main high state to identify anomalous low states (ALSs) with an objective and reproducible criterion. The method recovers the known ALSs, identifies two previously unrecognized candidate episodes, ALS 1991 and ALS 2015, and independently recovers the recently reported ALS 2025. We then characterized the long-term superorbital phase evolution, identifying seven epochs that alternate between smooth and variable behavior. The smooth component is empirically described by a cubic trend, while the recurrent variable epochs have a characteristic timescale of approximately 3400 days. Within a first-order tilted-disk framework, the secular phase evolution corresponds to an inferred cumulative disk-mass increase of $\sim1.8\%$ and an outer-disk-radius expansion of $\sim0.9\%$, while the short-term phase excursions can be reproduced by disk-mass variations of $\lesssim2.4\%$. A viscous diffusion timescale of approximately 168 days provides a characteristic timescale for the short-term disk response. Finally, the stable phase relation and high profile similarity between the main and short high states indicate that the warped disk maintains a remarkably stable global geometry over the observational baseline.

\end{abstract}
%%\keywords{TBD}


\section{Introduction} \label{intro}

%Basic property

First identified by the Uhuru satellite \citep{tan72}, Her X-1 is an eclipsing X-ray binary with a 1.7-day orbital period, consisting of an accreting X-ray pulsar with a 1.24-s pulsation period. The binary system comprises a 1.5 $M_{\sun}$ neutron star \citep{rey97} and a 2.35-2.45 $M_{\sun}$ companion star~\citep{lea14}. In addition to its orbital and spin periodicities, the system exhibits a $\sim$35-day superorbital modulation in the X-ray band. This long-term variation was first discovered by~\citet{tan72}, who reported a $35.7 \pm 0.3$-day intensity modulation from the Uhuru light curve, characterized by a high state lasting $\sim$9 days and a weak or undetectable state for the remaining $\sim$27 days. A subsequent analysis of the Uhuru data by~\citet{gia73} refined this superorbital period to $34.88 \pm 0.12$ days, describing a profile with an on-state of 11-12~days followed by a 24-day off-state. They also discovered that the turn-on (TO) times of the high state occur preferentially at orbital phases of either 0.2 or 0.7, a pattern later shown by~\citet{dav74} to be quasi-periodic rather than strictly regular. Using X-ray detectors aboard the Copernicus satellite,~\citet{fab73} identified a secondary, weaker emission peak within the superorbital cycle, termed the short high state (SHS). Its peak flux is lower than or comparable to that of the primary peak, known as the main high state (MHS). Taking the TO time as the fiducial point, the standard superorbital modulation profile consists of the MHS covering phases 0 to 0.31, the SHS covering phases 0.57 to 0.79, and low states with near-zero intensity during the remaining phases~\citep{sco99}. Furthermore, pre-eclipse X-ray dips are observed with a period of 1.65~days, which represents the beat period between the orbital and superorbital cycles~\citep{gia73, cro80, sco99}. During the low and eclipsing states, X-ray observations~\citep{cho94, lea95, jim05, lea15} reveal that the system features an ionized accretion disk atmosphere and corona, where electrons scatter the hard X-rays emitted from the neutron star.

To explain the 35-day superorbital modulation of Her X-1,~\citet{kat73} suggested that it results from occultation by a precessing ring at the outer rim of the accretion disk surrounding the neutron star. Furthermore,~\citet{pet75} proposed a model where Her X-1 hosts a twisted/warped disk precessing with a period of $\sim$35 days under tidal interactions from the companion star, which periodically obscures the central X-rays. Conversely,~\citet{rob74},~\citet{ger76}, and~\citet{lev82} suggested that a precessing tilted disk is the structure responsible for this 35-day superorbital modulation. Despite the success of these early geometric models in explaining the observed modulation, sustaining such a tilted disk against tidal dissipation remains a theoretical challenge. In a realistic fluid disk, differential tidal torques and internal viscous dissipation naturally damp out the disk tilt, forcing it to realign with the binary orbital plane~\citep{lub00, bat00}. To address this difficulty, the theory of radiation-driven warping was introduced \citep{pet77, pri96}, wherein radiation pressure from the central source exerts a feedback torque on the warped outer disk, allowing a small initial tilt to grow exponentially. Applying this mechanism to a full fluid-dynamical framework, \citet{ogi01} demonstrated that while radiation-driven warping is generally unstable in most X-ray binaries, stable solutions yielding a steady, rigidly precessing disk shape exist within a narrow range of parameters close to the stability limit. Crucially, Her X-1 was found to lie precisely within this stable regime, theoretically allowing the disk to maintain a highly stable, coherent shape over long timescales.

Unlike other sources exhibiting superorbital modulation, Her X-1 uniquely features anomalous low states (ALSs), during which both the X-ray flux and the superorbital profile become severely depressed or even undetectable. Several such episodes have been reported since the discovery of the source. The first event (ALS 1983) was documented by~\citet{par85}, who noted that the European X-ray Observatory SATellite (EXOSAT) failed to detect the expected superorbital modulation between late June and late August 1983, observing only low-level X-ray emission. Optical observations conducted prior to ALS 1983 (from June 1 to 17, 1983) revealed that the 35-day optical variability was gradually decaying, whereas the X-ray irradiation of HZ Her remained strong~\citep{del83}. This decoupled behavior implies that the X-ray emission remained intrinsically strong but was obscured by a tilted, precessing accretion disk~\citep{par85}. The second event (ALS 1993) was identified by~\citet{vrt94} through multi-wavelength observations of Her X-1 between August 10 and 14, 1993, which revealed X-ray fluxes at only 30\% to 50\% of the normal level; this state persisted for three to four superorbital cycles~\citep{vrt96}. Consequently,~\citet{vrt96} formally defined an ALS as a substantial, unexpected drop in X-ray flux persisting for at least two superorbital cycles, characterized by no significant change in the absorbing column density, minimal variation in the UV and optical flux, a spin-down trend (increase in pulse period), and an absence of pulsed emission above $\sim$1.0 keV.

The third ALS (ALS 1999) was first discovered by~\citet{lev99}, as the All-Sky Monitor (ASM) aboard the Rossi X-Ray Timing Explorer (RXTE) failed to detect the expected MHS around March 3, 1999. The upper limit of the flux was $\sim$15 mCrab, which is substantially lower than the typical 40-100 mCrab range. Subsequently, a number of follow-up X-ray and optical observations were conducted~\citep[e.g.,][]{par99,mar99,cob00,sti01}. Optical observations during ALS 1999 revealed that HZ Her was still strongly irradiated by the central X-ray source. Throughout most of ALS 1999, the superorbital modulation profile was completely diminished, leaving only a very low residual RXTE/ASM count rate \citep[see, e.g., Figure 1 in][or Figure 4 in~\citealp{kot12}]{sti01}.~\citet{oos01} reported that the first MHS following the cessation of ALS 1999 commenced on October 9, 2000, indicating that this anomalous state persisted for more than 586 days ($\sim$17 cycles). To date, ALS 1999 remains the episode with the longest documented duration. The subsequent ALS (ALS 2003) began around August 15, 2003 \citep{sti04} and ended on approximately July 11, 2004~\citep{boy04}. Similar to ALS 1999, the superorbital modulation profile was undetectable for most of the duration of ALS 2003~\citep[see, e.g., Figure 1 in][or Figure 4 in~\citealp{kot12}]{sti01}, though it exhibited a shorter duration of approximately 9 supercycles. The latest ALS, active near the end of 2025 (ALS 2025), was reported by~\citet{mih26}.  Based on the fact that the X-ray irradiation of the companion star from the central accreting neutron star remains constant during ALSs~\citep{del83,mar99}, these states are widely considered to be caused by the partial or total occultation of the central X-rays. This obscuration is driven either by a tilted or warped disk resulting from changes in the disk inclination angle \citep[e.g.,][]{cob00,cla03} or by an increased twist or warp of the accretion disk \citep{lea10a}.

Over the decades, numerous observational studies have addressed the stability of the superorbital modulation in Her X-1.~\citet{dav74} initially suggested that the superorbital period varies systematically on a quasi-period of $\sim$560 days. In contrast, by applying a dynamic power spectrum analysis to the light curve from the RXTE ASM,~\citet{cla03} demonstrated that the superorbital period remained essentially stable throughout the $\sim$7-year observational span. This long-term stability was subsequently corroborated by~\citet{kot12}, who analyzed $\sim$15 years of RXTE/ASM data using the same methodology. Interestingly, however, the dynamic folded light curve constructed by~\citet{cla03} revealed a pronounced phase drift of $\sim$0.1 supercycles, emerging approximately 15 supercycles prior to the onset of the ALS 1999. Analyzing the superorbital phase evolution allows for a more precise probing of the variation in the superorbital period. Taking the mean time of the MHS as the fiducial point,~\citet{oos01} analyzed the light curves collected by the RXTE ASM from January 1996 to February 2001. They found that, in addition to the phase variation prior to ALS 1999, a significant phase difference of $\sim$0.2 cycles existed before and after ALS 1999. Similarly, by selecting the mid-point of the MHS as the fiducial point,~\citet{sti04} reported the superorbital phase evolution through an analysis of data collected by the Burst and Transient Source Experiment (BATSE) aboard the Compton Gamma Ray Observatory (CGRO) and the RXTE ASM up to MJD $\sim$53000. They suggested that the dataset can be divided into four distinct epochs separated by ALSs, with each epoch characterized by its own ephemeris. Furthermore, by compiling the TO times spanning from 1971 to 2012, \citet{sta06,sta09,sta13} proposed that Her X-1 likely exhibits a quasi-periodic superorbital phase variation of $\sim$5 or $\sim$10 years, which is potentially driven by the magnetic activity cycle of the companion star~\citep{sta09}.

In this study, we investigate the long-term variation of the superorbital modulation in Her X-1 using data collected by sky survey and monitoring telescopes, which have continuously monitored the source with sufficient sensitivity from 1991 to 2026. A brief overview of these instruments, including the CGRO BATSE, RXTE ASM, the Burst Alert Telescope aboard the Swift Observatory (Swift BAT), MAXI  Gas Slit Camera (MAXI GSC), Fermi Gamma-ray Burst Monitor (Fermi GBM), is presented in Section~\ref{obs}. The initial data reduction, light curve folding, modeling of the modulation profiles, and the selection of fiducial points are described in Section~\ref{mp}. In Section~\ref{ials}, we introduce a newly developed methodology to identify ALSs from three of the five light curves analyzed in this work, and we report two newly identified candidates and independently recovered ALS 2025. The superorbital phase evolution of Her X-1 spanning a $\sim$35-year baseline is demonstrated in Section~\ref{spe}, encompassing both long-term evolutionary trends and short-term variations. In Section~\ref{compms}, we compare the properties of the MHS and SHS, including the distribution of their phase differences and the similarity of their modulation profiles. The investigation of a possible energy-dependent superorbital phase lag between soft (RXTE ASM and MAXI GSC) and hard X-ray bands (CGRO BATSE and Swift BAT) is also presented in this section. Finally, Section~\ref{dis} discusses the physical interpretations derived from our data analysis, which include the geometric disk configuration during ALSs (Section~\ref{dis_als}), the plausible physical mechanisms driving the observed superorbital phase variations and evolution (Section~\ref{dis_spe}), and the structural stability of the accretion disk (Section~\ref{dis_stdisk}).

%%the implications of our analysis and comment on the possible scenarios for the negative measured orbital period derivative. 

\section{Observations} \label{obs}

The CGRO~\citep{geh93}, active from 1991 to 2000, featured the BATSE to detect gamma-ray bursts (GRBs) and high-energy transients. Eight corner-mounted modules provided near-complete sky coverage, each pairing a spectroscopy detector with a Large Area Detector (LAD) utilizing 50.8 cm diameter NaI(Tl) scintillators optimized for the 20 keV-1 MeV range ~\citep{fis94}. BATSE's all-sky monitoring demonstrated that GRBs are distributed isotropically, confirming their cosmological origin~\citep{mee92}. For this study, we analyzed the 20-40 keV light curve of Her X-1 from the NSSTC Gamma-Ray Astrophysics Team archive\footnote{ \url{https://gammaray.nsstc.nasa.gov/batse/occultation/sourcepages/herx-1_lc.html}}, spanning May 30, 1991 (MJD 48406) to May 26, 2000 (MJD 51690).

The RXTE ASM~\citep{lev96} tracked variable and transient X-ray sources using three scanning shadow cameras with position-sensitive proportional counters and a one-dimensional coded mask, yielding a $6^{\circ} \times 90^{\circ}$ field of view. The ASM scanned nearly the entire sky every 90 minutes across 1.5-12 keV, telemetry-split into 1.5-3 keV, 3-5 keV, and 5-12 keV sub-bands, producing both 90-second "dwell" exposures and daily averages. We retrieved the 1.5-12 keV Her X-1 dwell light curve from the HEASARC archive. Adhering to the data selection criteria of~\citet{lev11}, to avoid late-mission background degradation from Al K\({}_{\alpha }\) photons, we restricted our superorbital modulation sample to the interval from January 5, 1996 (MJD 50087) to January 28, 2010 (MJD 55224).

The Burst Alert Telescope (BAT) aboard the Neil Gehrels Swift Observatory (hereafter, Swift) is a 1.4-steradian field-of-view, coded-mask instrument monitoring the 15-150 keV hard X-ray sky since 2004 \citep{bar05}. Optimized for GRB localization, its 5200 cm\({}^{2}\) collecting area and 20\arcmin\ angular resolution also enable continuous tracking of known X-ray sources during Swift's ~96-minute orbits, facilitating long-term variability studies. We analyzed the 15-50 keV light curve of Her X-1 from the Swift monitoring archive\footnote{\url{https://swift.gsfc.nasa.gov/results/transients/HerX-1/}}, covering February 15, 2005 (MJD 53416) to January 28, 2026 (MJD 61068).

The Monitor of All-sky X-ray Image (MAXI) on the International Space Station (ISS) detects X-ray transients and tracks long-term source variability \citep{mat09}. MAXI utilizes a Gas Slit Camera (GSC; 5250 cm\({}^{2}\) effective area, 2-30 keV) and a Solid-state Slit Camera (SSC; 200 cm\({}^{2}\) effective area, 0.5-12 keV), scanning almost the full sky twice per ~90-minute ISS orbit. To study the superorbital modulation of Her X-1, we used the 2-10 keV GSC light curve from the MAXI on-demand web service\footnote{\url{http://maxi.riken.jp/mxondem}}, spanning August 12, 2009 (MJD 55055) to January 28, 2026 (MJD 61068).

The Fermi Gamma-ray Burst Monitor (GBM), aboard Fermi since 2008, tracks GRBs and transients across 8 keV to 40 MeV using the Earth occultation technique \citep{mee09, wil12}. It features 12 low-energy NaI detectors and two high-energy BGO detectors arranged for nearly complete sky coverage to facilitate prompt multi-wavelength follow-up. For this work, we examined the 12-50 keV light curve of Her X-1 hosted on the NSSTC website\footnote{\url{https://gammaray.nsstc.nasa.gov/gbm/science/earth_occ/HERX-1.html}}, encompassing the earliest available data from February 13, 2012 (MJD 55970) through January 28, 2026 (MJD 61068).

\begin{figure}
  \epsscale{1.25}
\plotone{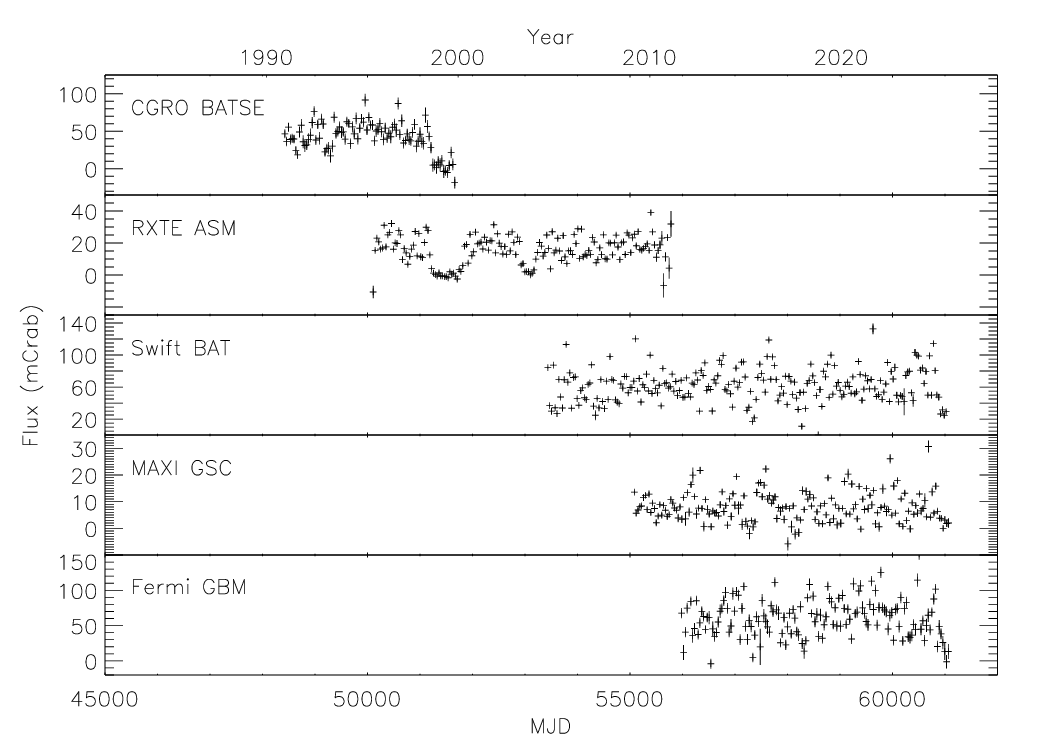}
\caption{Light curves collected by five instruments for analysis in this study. The bin size of these light curves is 34.7937 days (about a superorbital cycle, see Section~\ref{mp}).  \label{lcurves}}
\end{figure}

The light curves of Her X-1 collected by these five instruments are shown in Figure~\ref{lcurves}.

\section{Data Analysis} \label{da}
\subsection{Modulation Profile and Selection of Fiducial Point}\label{mp}

\begin{figure}
  \epsscale{1.2}
\plotone{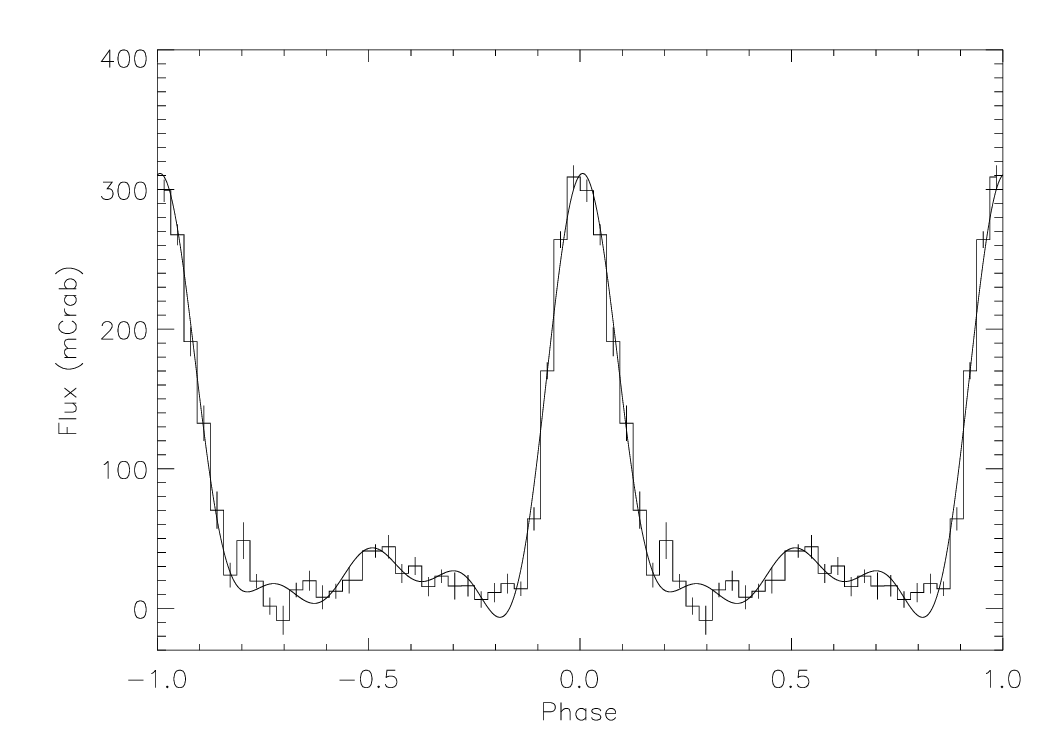}
\caption{A representative superorbital modulation profile constructed from Swift BAT observations between MJD 53822.067 and 53856.861 (one supercycle). The solid line represents the best-fit four-component sinusoidal model. \label{typ_prof1}}
\end{figure}

To accurately evaluate the superorbital modulation profile, data points collected during eclipsing states were excluded from the analysis. All light curve timestamps were first converted to the Solar System Barycenter (SSB)  using the {\it earth2sun} task in {\it HEASOFT}. Based on the latest eclipsing ephemeris provided in Table 2 of \citet{yan26}, data within the orbital phase range of -0.25 to 0.1 were rejected. A folding period of $p_f=$34.7937 days was derived from the mean period across the Lomb-Scargle periodograms \citep{sca82} of the five light curves. Using this period and the midpoint of the entire observation span (MJD 54737.0) as the initial reference epoch, the superorbital modulation profile was clearly resolved, even for data segments spanning only a single superorbital cycle. We modeled this modulation profile using a multicomponent sinusoidal function:

\begin{equation}\label{multisinu}
  r(\phi) = a_0 + \sum_{k=1}^{n} \left[a_k \cos(2\pi k \phi) + b_k \sin(2\pi k \phi)\right]
\end{equation}

\noindent where $r$ represents the observed flux, $\phi$ is the superorbital phase, and $n$ denotes the number of sinusoidal harmonics. We determined that the modulation profile is well-described by a four-component sinusoidal function (n = 4). A representative superorbital modulation profile alongside its optimal model fit is displayed in Figure \ref{typ_prof1}.

Because defining a precise TO time can be challenging in light curves obtained from sky surveys and monitoring telescopes, following~\citet{lea20}, we  selected the MHS peak of the best-fit four-component sinusoidal model as our fiducial point. Utilizing the Swift BAT data taken from several supercycles around the midpoint (MJD 54737.0), we slightly adjusted the zero-phase epoch to align the fiducial point closely with zero, establishing the following linear ephemeris:

\begin{equation}\label{lieph} 
T_N = \text{MJD } 54726.703 + 34.7937N 
\end{equation}

\noindent This ephemeris was strictly applied to all subsequent analyses. The statistical uncertainties for the parameters of the modulation profile, such as the phase alignments and the peak flux heights of the MHS and SHS were robustly evaluated through $10^4$ Monte Carlo realizations.

\subsection{Identification of New Anomalous Low State Candidates }\label{ials}

Since the discovery of Her X-1, five individual ALS episodes have been reported (see Section~\ref{intro} for a detailed review). However, based on the criteria established by~\citet{vrt96}, a rigid flux threshold separating the normal state from an ALS remains undefined. In this work, we presented an objective methodology to constrain this threshold by analyzing the MHS peak heights extracted from long-term monitoring and all-sky survey light curves.

We selected the light curves from CGRO BATSE, RXTE ASM, and Swift BAT because their combined timeline covers the entire duration of this study (MJD 48406 to 61068) with superior sensitivity. Since high time resolution is critical for this method, we divided the light curves into 1-supercycle segments based on Eq.~\ref{lieph} and folded them using the same ephemeris to obtain 32-bin folded light curves. Folded light curves with excessive missing data points within the expected MHS phase range were excluded from the analysis, as the heights of their MHS could not be accurately evaluated. The remaining folded light curves were then fitted using the 4-component sinusoidal model described in Eq.~\ref{multisinu}. The maximum values of the optimally fitted model were taken as the MHS heights, even in cases where the superorbital modulation profiles were completely diminished, as observed in ALS 1999 and 2003.

\begin{figure}[htbp]
  \centering
  \includegraphics[width=0.45\textwidth]{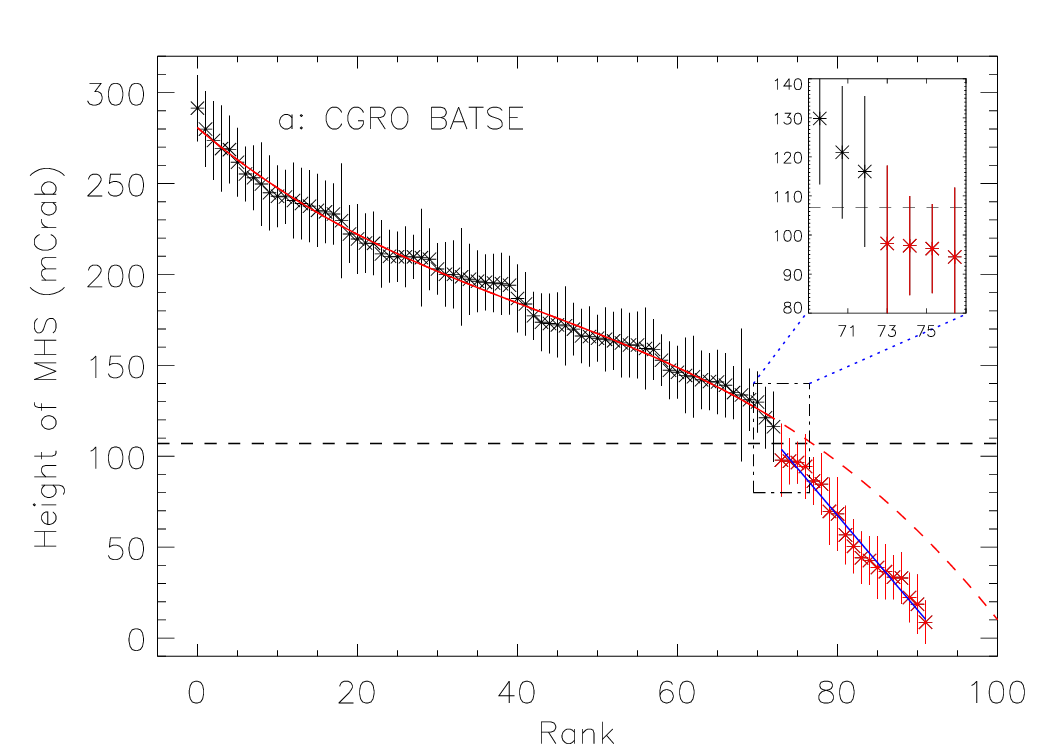}
   \includegraphics[width=0.45\textwidth]{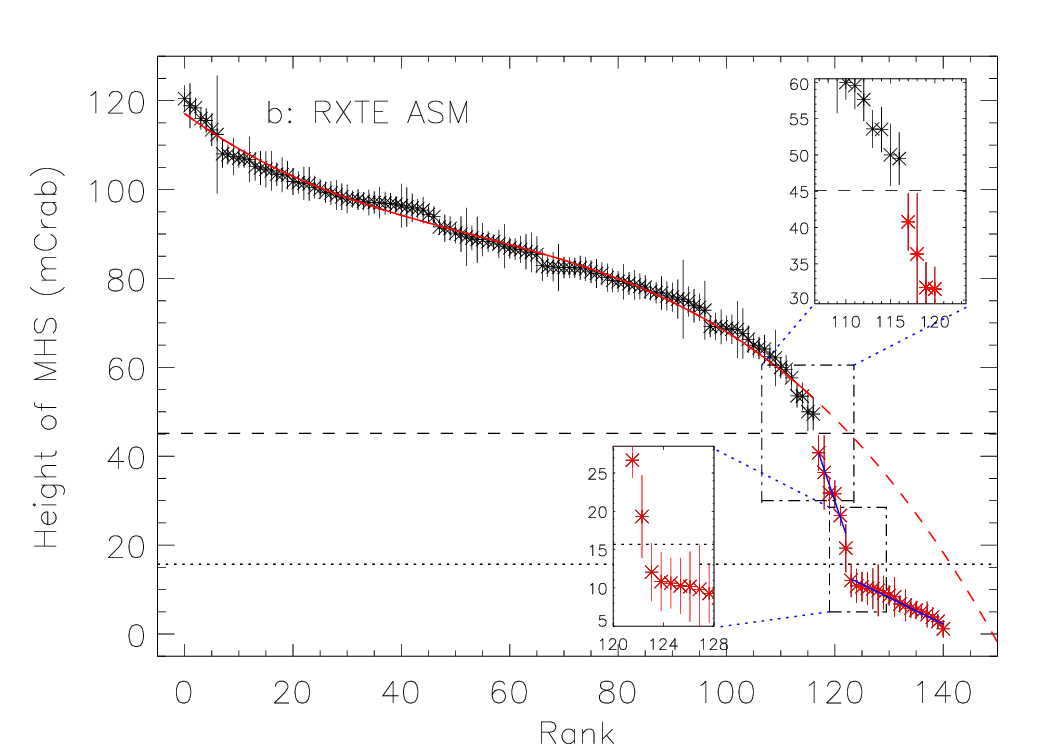}
 \includegraphics[width=0.45\textwidth]{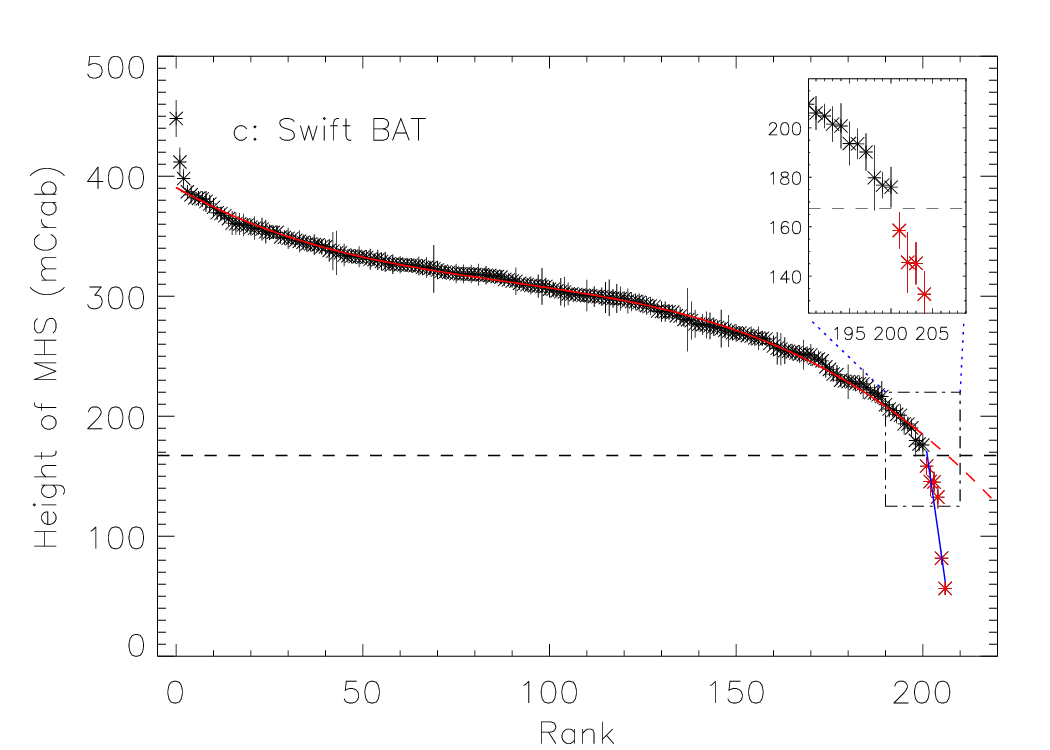}
\caption{The rank spectra used to identify ALSs, constructed from observations with (a) CGRO BATSE, (b) RXTE ASM, and (c) Swift BAT. The gaps are clearly visible in the zoomed-in panels. The dashed horizontal lines indicate the thresholds between the normal and transition states, whereas the dotted horizontal line in panel (b) marks the threshold between the transition and deep low states. Data points with MHS heights below the dashed thresholds are highlighted in red and represent low-state measurements associated with the identification of ALS intervals. The red curves show the empirical trends of the normal-state rank spectra, modeled with cubic functions to correctly model the tilts near the beginning and end, while the red dotted curves represent extrapolations of these trends. The blue lines indicate the empirical linear  trends of the low-state data, which clearly deviate from the normal-state trends.  \label{rank}}
\end{figure}

For each light curve, we sorted the MHS heights in descending order and assigned a rank to each value. Figure~\ref{rank} displays the evolution of MHS heights as a function of rank number. Interestingly, at least one gap, defined as a significant difference in MHS height between two consecutive ranks, can be observed, with two distinct gaps identified in the RXTE ASM data. Furthermore, the trend changes clearly below these gaps. Above the gap (or above the first gap for RXTE ASM), the trend is approximately linear, though it tilts near the beginning and end. Below the gap, the behavior shifts, indicating a transition in the state of Her X-1 as it enters an ALS. To define the duration of these ALS intervals, we established the state-change threshold as the mean MHS height between the two consecutive ranks defining the gap. The statistical significance of these gaps and their corresponding thresholds are summarized in Table~\ref{threshold}.

\begin{table}
\begin{center}
  \caption{Significance of the Gap and the Threshold for ALSs Identified in This Study\label{threshold}}

  \begin{tabular}{cccc}
    \\
    \tableline\tableline
     \tableline\tableline
Instrument   & Gap     & Significance\tablenotemark{a} & Threshold\\
             &         &        & (mCrab)    \\ 
\tableline
CGRO BATSE   &  1      & 4.18        & 18.42        \\
\tableline
RXTE ASM    &  1      & 5.56        & 8.70       \\
            &  2      & 4.53        &  7.25        \\
\tableline
Swift BAT   &  1      & 7.62        &  17.53       \\

\tableline\tableline
\tableline\tableline
  \end{tabular}
      \tablenotetext{}{\parbox[t]{0.8\columnwidth}{\raggedright \bf Notes:}}
  \tablenotetext{a}{\parbox[t]{0.8\columnwidth} {\raggedright The significance is defined as $|\Delta h_{gap} - \bar{\Delta h}| / {\sigma_{\Delta h}}$, where $\Delta h_{gap}$ is the height difference of the gap, $\bar {\Delta h}$ and ${\sigma_{\Delta h}}$ are the mean and standard derivation of the height difference between two consecutive ranks}}
  
\end{center}

\end{table}

\begin{figure}[htbp]
  \centering
  \includegraphics[width=0.45\textwidth]{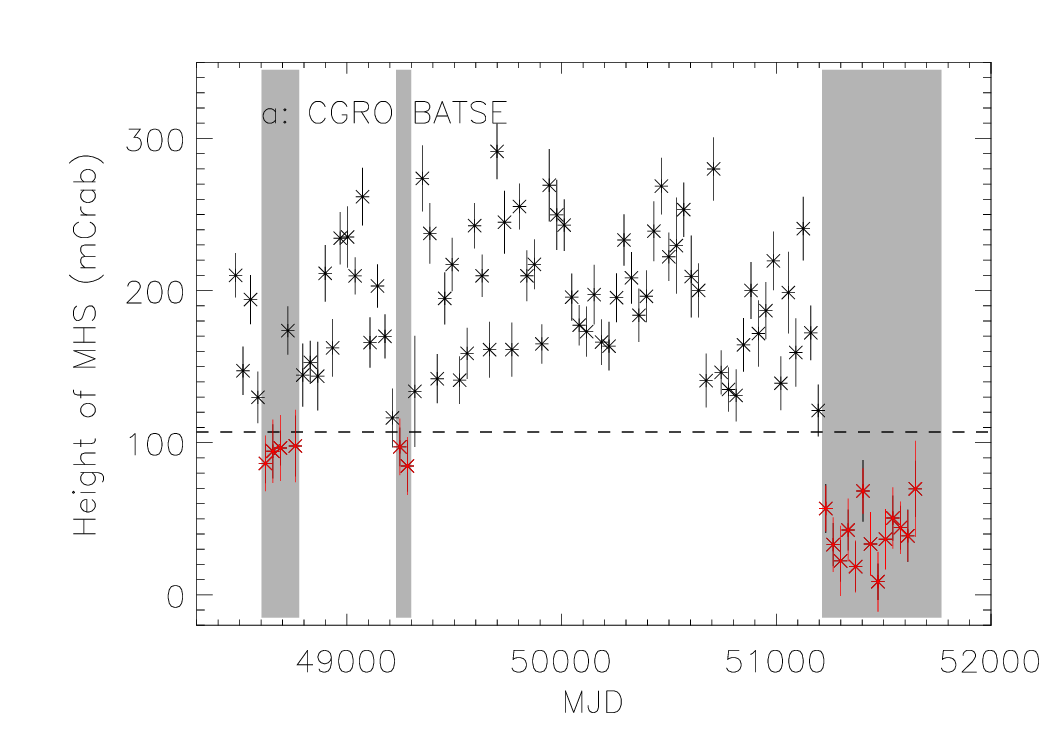}
   \includegraphics[width=0.45\textwidth]{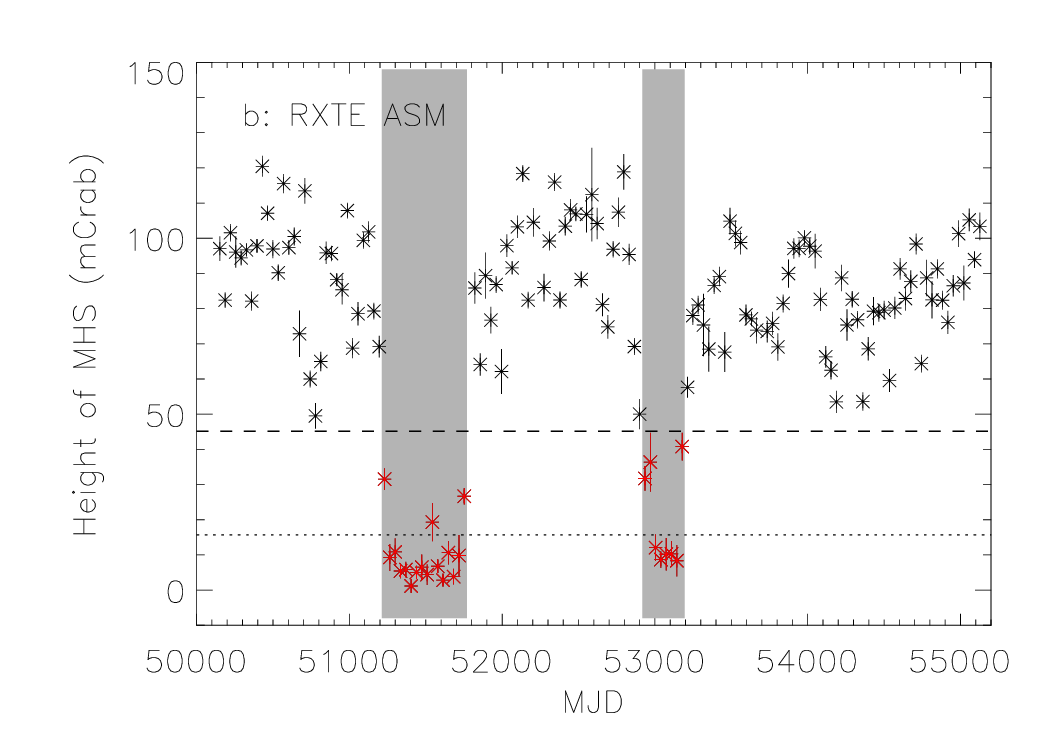}
 \includegraphics[width=0.45\textwidth]{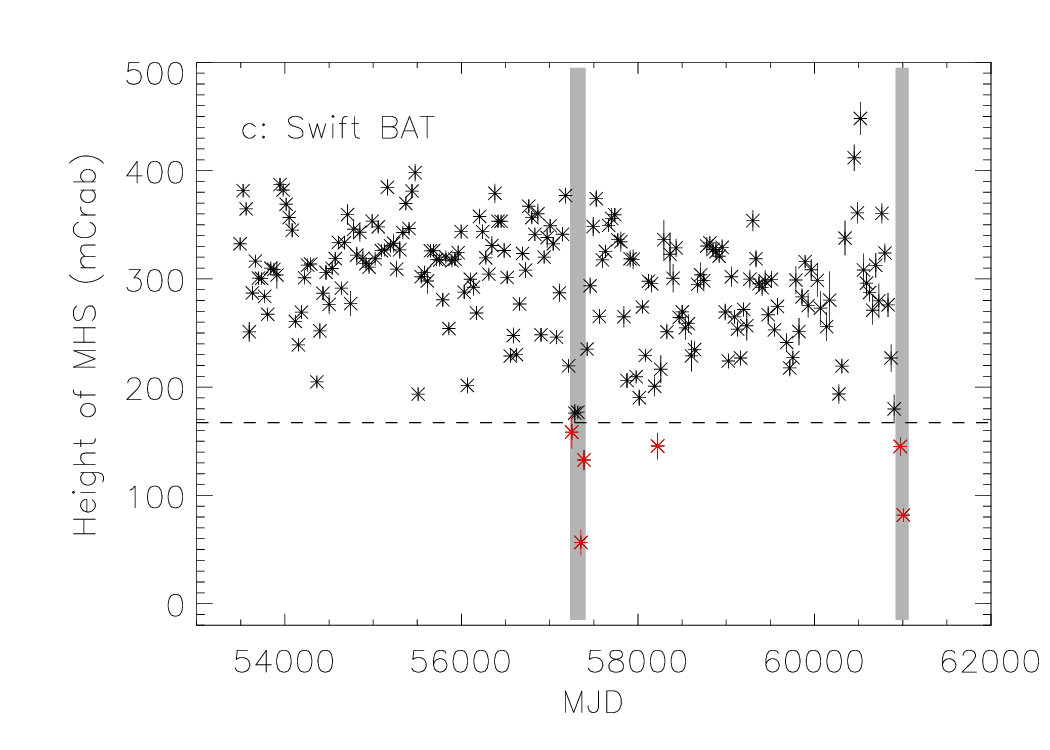}
\caption{The evolution of the MHS heights, constructed from observations with (a) CGRO BATSE, (b) RXTE ASM, and (c) Swift BAT. The dashed horizontal lines indicate the thresholds between the normal and transition states, whereas the dotted horizontal line in panel (b) marks the threshold between the transition and deep low states. Data points with MHS heights below the dashed thresholds are highlighted in red. The midpoints of the transitions between the normal and transition states are adopted as the boundaries of the corresponding low-state intervals. Following the duration criterion of ~\citet{vrt96}, newly identified low states persisting for at least two superorbital cycles are classified as ALS candidates. The gray shaded regions mark the ALS intervals identified in this study, including both previously reported ALSs and newly identified candidates (see Table~\ref{als_tb}). Consequently, the isolated low state observed around MJD 58223, which lasted for only one superorbital cycle, is not classified as an ALS.\label{rank_select_mh}}
\end{figure}

We applied these thresholds to the MHS height vs time plots, as shown in Figure~\ref{rank_select_mh}. In the CGRO BATSE observations, besides identifying the previously reported ALS 1993~\citep{vrt94} and 1999~\citep{lev99}, we also found a cluster of consecutive data points between MJD 48600 and 48800 (from late 1991 to early 1992) with the MHS heights below the threshold. This interval lasted for approximately 5 supercycles, except for a single point that jumped above it for unknown reasons. According to the intensity-based definition of an ALS proposed by~\citet{vrt96}, an unexpected drop in X-ray flux persisting for at least two superorbital cycles, Her X-1 may have also entered an ALS (hereafter ALS 1991) during this timeframe, as seen in the top panel of Figure~\ref{rank_select_mh}. However, because other diagnostic criteria of an ALS (such as the non-detection of pulsations above $\sim$1.0 keV) could not be verified, we classify this ALS 1991 event as an ALS candidate.

\begin{figure}[htbp]
  \centering
  \includegraphics[width=0.48\textwidth]{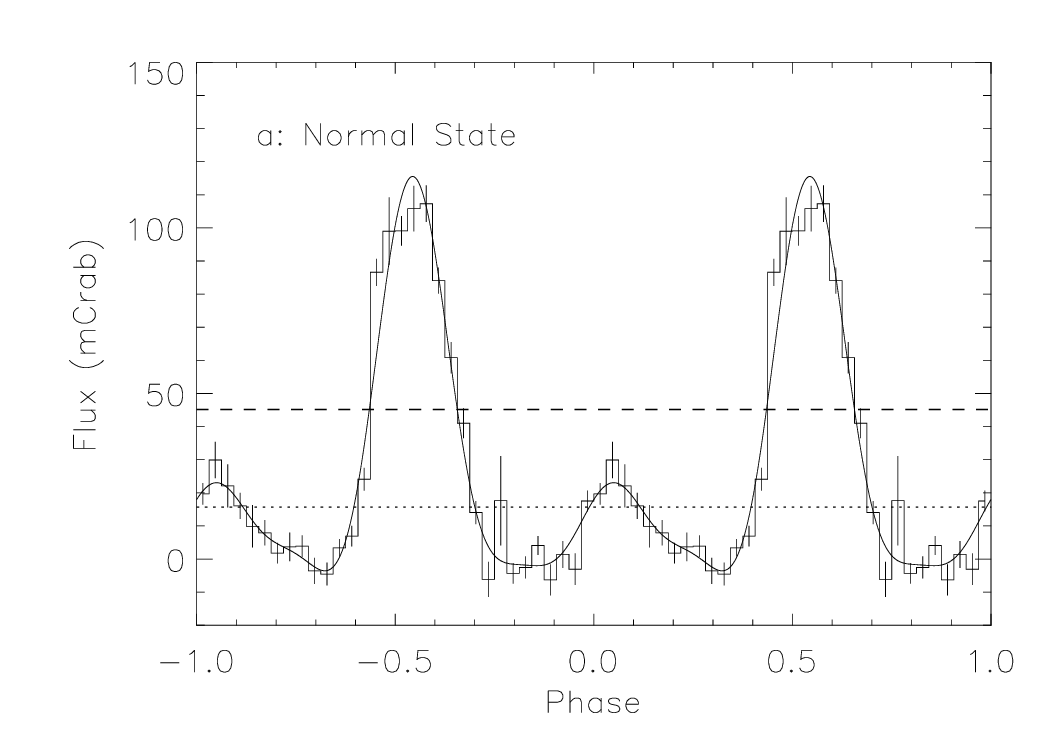}
   \includegraphics[width=0.48\textwidth]{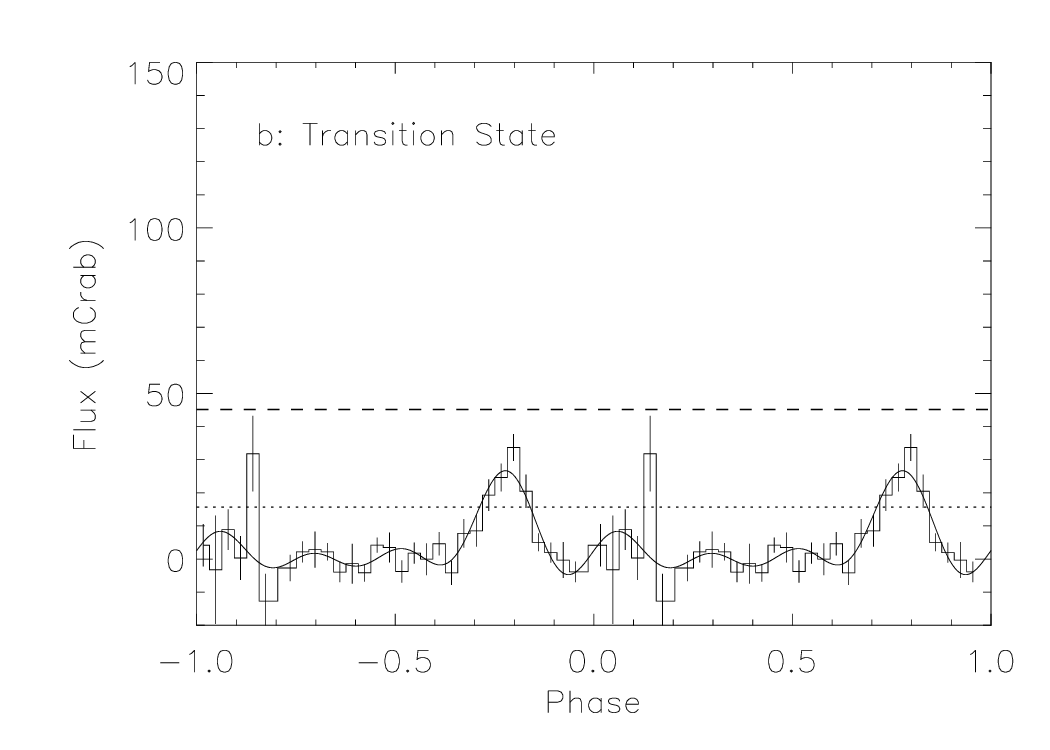}
 \includegraphics[width=0.48\textwidth]{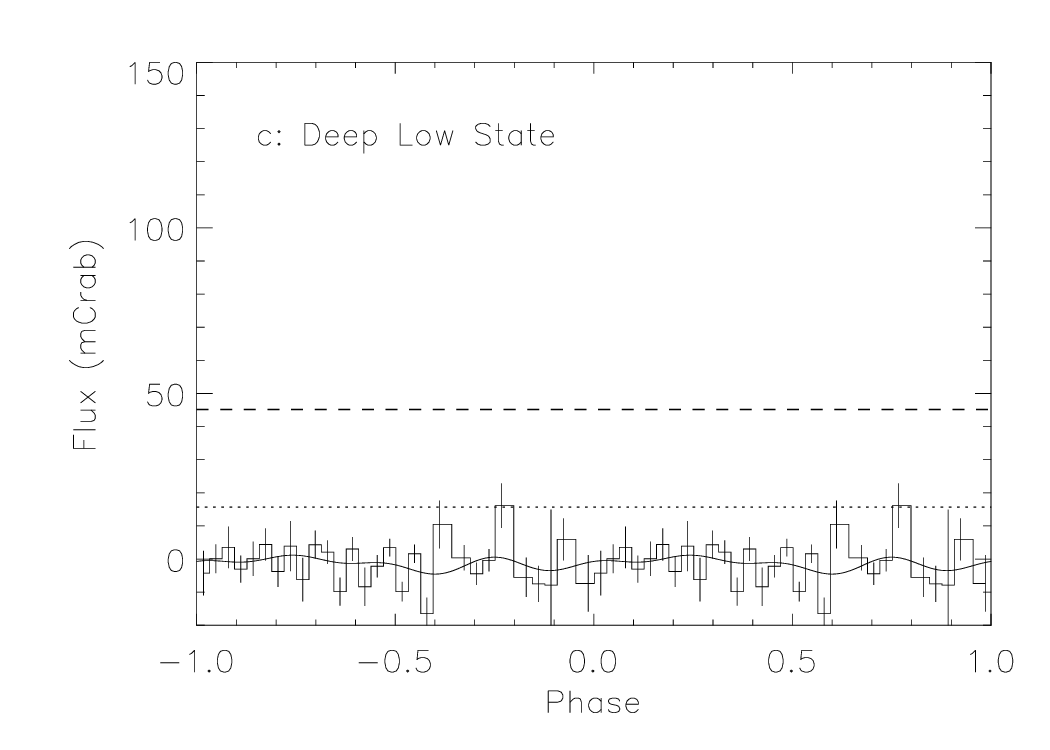}
\caption{Typical superorbital modulation profiles of Her X-1 constructed from RXTE ASM observations in different states. (a) Normal state, in which both the MHS and SHS are significantly detected. (b) Transition state, in which the MHS height falls below the normal-to-transition threshold (dashed line) but remains detectable, whereas the SHS is nearly undetectable. (c) Deep low state, in which the MHS height falls below the transition-to-deep-low threshold (dotted line) and no significant superorbital modulation profile is detectable. \label{typ_prof3}}
\end{figure}

In the RXTE ASM observations (see the middle panel of Figure~\ref{rank_select_mh}), both long-lasting ALSs, ALS 1999 and 2003~\citep{sti04}, were clearly identified. According to the thresholds defined in this study, Her X-1 first entered a {\it transition state} for 1 to 2 supercycles, with MHS heights ranging between 8.70 and 7.25 mCrab (Thresholds 1 and 2, as listed in Table~\ref{threshold}), before plunging into a {\it deep low state}\footnote{Physically, the superorbital modulation is almost undetectable in this state, meaning that the true MHS height is basically undefined. The so-called 'MHS height' derived for this state in this study simply represents the maximum value of the best-fit model, which inherently yields an expectedly low value.} where the MHS height became virtually undetectable. Near the end of these long-lasting ALSs, Her X-1 first returned to the transition state before fully recovering to the normal state. The typical modulation profiles for the normal, transition, and deep low states are displayed in Figure~\ref{typ_prof3}. As noted in the middle and bottom panels of Figure~\ref{typ_prof3}, the MHS remains marginally detectable during the transition state, whereas the superorbital modulation profile completely disappears in the deep low state. Notably, while ALS 1999 was also identified in the CGRO BATSE data, no clear gap between the transition and deep low states could be resolved via rank selection (see the top panel of Figure~\ref{rank}), which is very likely due to the lower sensitivity of CGRO BATSE. Given the lack of a resolved second gap in the CGRO BATSE data, ALS 1991 and ALS 1993 are classified here as transition-state events; however, the lower BATSE sensitivity prevents us from ruling out a deeper low state.

By applying the threshold of 17.53 mCrab derived from the Swift BAT observations to the MHS height vs time plot (see the bottom panel of Figure~\ref{rank_select_mh}), a new ALS candidate spanning approximately MJD 57200 to 57500, hereafter ALS 2015, was identified. This ALS lasted for about 5 supercycles. Similar to ALS 1991, two data points jumped above the threshold, though they remained very close to it, which can likely be attributed to the uncertainties in the MHS heights. Thus, as with ALS 1991 and 1993, we suggest that Her X-1 remained in the transition state during this ALS. Furthermore, a single supercycle with an MHS height below the threshold was detected around MJD 58200. However, because this low state lasted for only one supercycle, it was not classified as an ALS according to the definition established by~\citet{vrt96}. Finally, the recently discovered and still ongoing ALS 2025~\citep{mih26} was also identified in this study (see the bottom panel of Figure~\ref{rank_select_mh}), although we are currently unable to determine whether it will reach the deep low state.

We therefore defined the ALS time intervals using the midpoints of Her X-1's transitions: from the normal to the transition state (start time), and from the transition back to the normal state (end time). All properties of the ALSs identified in this study, including their start and end times, durations, and types (transition state only, or transition plus deep low state), are summarized in Table~\ref{als_tb}. Table~\ref{als_comp} compares the historically reported start and end times of ALS 1993, 1999, and 2003 with those obtained in this work. In general, the differences are $\lesssim$ 1 supercycle, with the likely exceptions of the end time of ALS 1999 and the start time of ALS 2003. A more detailed discussion on the validity of this rank-selection method is presented in Section~\ref{dis_als}.

\begin{table*}[ht]
\centering
  \caption{Anomalous Low States Identified in This Study\label{als_tb}}

  \begin{tabular}{ccccccc}
    \\
    \tableline\tableline
     \tableline\tableline
ALS   & Year      & Start time & Stop time & Duration & Duration       & Type \\
index &           & (MJD)      & (MJD)     & (days)    & (superorbital cycle)  &  \\ 
\tableline
1991\tablenotemark{a}  & 1991/1992 & 48603      & 48777     &  174      & 5          &  T\tablenotemark{b}\\
1993    & 1993      & 49229      & 49299     &  70       & 2           &T  \\
1999    & 1999/2000 & 51213      & 51769     &  556      & 16          & T+D\tablenotemark{c} \\
 2003   & 2003/2004 & 52917      & 53195     &  278      & 8          & T+D \\
 2015\tablenotemark{a}    & 2015/2016 & 57232      & 57406     &  174      & 5          & T \\
2025\tablenotemark{a}    & 2025/2026 & 60920      & $>$61068  &  $>$148   & $>$4       & T+? \\
\tableline\tableline
\tableline\tableline
  \end{tabular}
      \tablenotetext{}{\parbox[t]{1.2\columnwidth}{\raggedright \bf Notes:}}
  \tablenotetext{a}{\parbox[t]{1.2\columnwidth}{\raggedright ALS candidates}}
  \tablenotetext{b}{\parbox[t]{1.2\columnwidth}{\raggedright Only transition (T) state is observed} }
  \tablenotetext{c}{\parbox[t]{1.2\columnwidth}{\raggedright Both transition (T) and deep low state (D), i.e. height of MHS $\approx$0}}

\end{table*}

\begin{table*}[ht]
\centering
  \caption{ALS Start and Stop Times Comparison \label{als_comp}}

  \begin{tabular}{cccccc}
    \\
    \tableline\tableline
     \tableline\tableline
ALS   & Start time & Start time         & Stop time & Stop time            & Reference  \\
    &   (MJD)    & difference (cycle)\tablenotemark{a}  & (MJD)     & difference (cycle)\tablenotemark{a}  &             \\ 
\tableline
1993    & 49213      & -0.45               &  49282    &  -0.48               &  \citet{sti04}          \\
     & 49229      &   -----               &  49299    &  -----                 &  This work  \\
\tableline
 1999   & 51195      & -0.50               & 51892     &  3.5                &   \citet{sti04}         \\
     & 51223      & 0.29                & 51831     &  1.7                 &  \citet{lea20}          \\
      & 51213     & -----                 & 51769     & -----                 &   This work  \\
 \tableline
 2003    & 52866     & -1.4                & 53197\tablenotemark{b}     & 0.05                &    \citet{sti04,boy04}           \\
      & 52915     & -0.06               & 53232     & 1.06                &   \citet{lea20}            \\
      & 52917     & -----                 & 53195     & -----                 &  This work  \\
\tableline\tableline
\tableline\tableline
  \end{tabular}
      \tablenotetext{}{\parbox[t]{1.78\columnwidth}{\raggedright \bf Notes:}}
  \tablenotetext{a}{\parbox[t]{1.75\columnwidth}{\raggedright Time differences between the corresponding times reported in references and this work}}
   \tablenotetext{b} {\parbox[t]{1.75\columnwidth}{\raggedright ALS ended before this MJD 53196.8~\citep[see][]{boy04}}}

\end{table*}

\subsection{Superorbital Phase Evolution}\label{spe}

To determine the superorbital phase more precisely, we segmented all five light curves into four-supercycle intervals based on Eq.~\ref{lieph}. We then folded these segments using the same linear ephemeris to produce 32-bin folded light curves. As described in Section~\ref{ials}, the phase of each folded light curve was defined by the peak position of a four-component sinusoidal model (Eq.~\ref{multisinu}), with the midpoint of the segment designated as the time tag. Because the modulation was weak during ALSs, all segments with time tags falling within an ALS were excluded from the analysis. However, because of the long duration of ALS 1999 (556 days; see Table~\ref{als_tb}), a cycle count ambiguity may arise~\citep[see][]{sta09,sta13}. When folding the light curves, the phase range was initially restricted to $[-0.5, 0.5]$. Thus, the initial phase of the data point immediately preceding ALS 1999 was 0.4303 (at MJD 51177.7456), whereas the one immediately following ALS 1999 was -0.2475 (at MJD 51873.6196). This yielded a phase difference of -0.6778 and a mean superorbital period of 33.3440 days during ALS 1999. However, if we shift the phase before ALS 1999 to 0.4303 - 1 = -0.5697, the phase difference becomes 0.3222. This corresponds to a mean superorbital period of 35.4828 days during ALS 1999, which is closer to $p_{f}=34.7937$ days (the mean superorbital period derived from power spectral analysis, see Section~\ref{mp}). We therefore selected the latter case and adjusted all preceding phases to ensure the continuity of the phase evolution.

\begin{figure}
  \epsscale{1.25}
\plotone{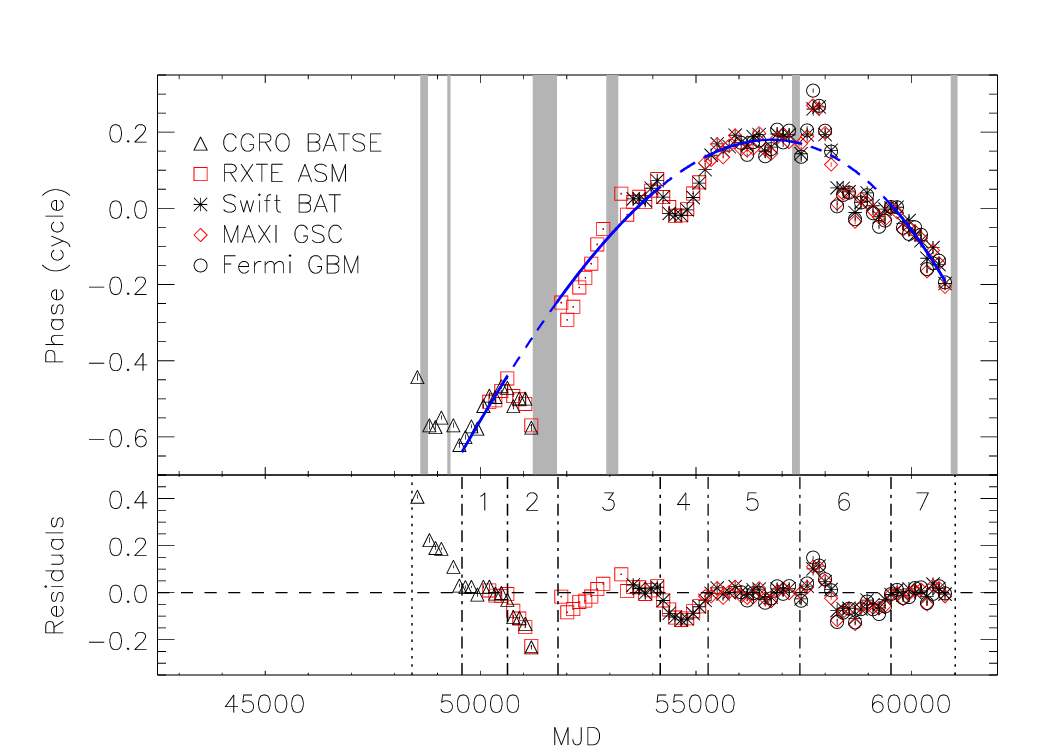}
\caption{Top: The superorbital phase evolution derived from the five X-ray light curves analyzed in this study. The solid blue lines represent the best-fit cubic trend determined from the data points in the smooth epochs, while the dashed blue lines show the continuation of this cubic trend through the variable epochs. The gray shaded regions mark the ALS intervals identified in this study. Bottom: The superorbital phase residuals after subtraction of the cubic trend. The numbers indicate the phase-variation epochs listed in Table~\ref{spve}. These epochs are separated by vertical dash-dotted lines, while the vertical dotted lines indicate the beginning and end of the data coverage. \label{phev}}
\end{figure}

The phase evolution after adjustment is shown in Figure~\ref{phev}. Tracing the phase variation from its minimum value at MJD 49507.6480 reveals a general concave-down trend superimposed with a recurrent ``smooth--variable'' pattern. In this pattern, the ``smooth'' phase follows a gradual evolutionary trend, whereas the ``variable'' phase deviates from it. We separated the phase evolution into smooth and variable types across seven epochs, as listed in Table~\ref{spve} and shown in Figure~\ref{phev}. We found that the individual epochs lasted approximately 1100 or 2200 days. The alternating smooth and variable behavior recurs on a timescale of several thousand days, although the phase evolution during Epoch 2 is only partially constrained because ALS 1999 spanned more than half of this epoch, during which the superorbital phase could not be reliably measured.

\begin{table*}[ht]
\centering
  \caption{Superorbital Phase Variation Epochs\label{spve}}

  \begin{tabular}{ccccccc}
    \\
    \tableline\tableline
     \tableline\tableline
Epoch & Start time & Stop time & Duration & Duration & Mean Deviation & Variation \\
index & (MJD)      & (MJD)     & (days)   & (years)  & (super cycle)  & Type \\ 
\tableline
1     & 49570      & 50627     & 1057     &  2.89    & 0.0181        &  S\tablenotemark{a}\\
2     & 50627      & 51796     & 1169     &  3.20    & 0.1519        &  V\tablenotemark{b}\\
3     & 51796      & 54170     & 2374     &  6.50    & 0.0367        &  S \\
4     & 54170      & 55283     & 1113     &  3.05    & 0.0831        &  V \\
5     & 55283      & 57411     & 2128     &  5.83    & 0.0194        &  S \\
6     & 57411      & 59528     & 2117     &  5.80   & 0.0766        &  V  \\
7     & 59528      & ---\tablenotemark{c}       & ---      &  ---    & ---        &  S \\
\tableline\tableline
\tableline\tableline
  \end{tabular}
      \tablenotetext{}{\parbox[t]{1.2\columnwidth}{\raggedright \bf Notes:}}
  \tablenotetext{a}{\parbox[t]{1.15\columnwidth} {\raggedright Smooth type: the superorbital phase evolution basically follows the cubic trend described in Section~\ref{spe} and Table~\ref{paracub}}}
  \tablenotetext{b}{\parbox[t]{1.15\columnwidth} {\raggedright Variable type: the superorbital phase evolution deviates substantially  from the cubic trend}}
\tablenotetext{c}{\parbox[t]{1.15\columnwidth} {\raggedright Because Her X-1 entered ALS 2025 on MJD 60920, the stop time of Epoch 7 remains undetermined (see Section~\ref{mag_per}). }}

\end{table*}

To assess the overall long-term trend of the superorbital phase evolution, we fitted polynomial models exclusively to the smooth epochs, excluding the variable epochs and the intervals affected by ALSs. An F-test comparing the quadratic and cubic models yielded an F-value of 39.11 with a p-value of $1.14\times10^{-8}$ ($N=99$), indicating that the cubic model provides a statistically significant improvement over the quadratic model. In comparison, adding a quartic term yielded an F-value of 6.67 with a p-value of $1.13\times10^{-2}$, indicating a statistically detectable additional improvement. Despite this improvement, we adopted the cubic polynomial as a more parsimonious empirical representation of the secular phase evolution. The best-fit parameters are listed in Table~\ref{paracub}. Importantly, the cubic function is not intended to imply that the underlying phase evolution is intrinsically governed by a cubic physical law; rather, it provides a convenient empirical description of the observed secular evolution over the time span of our data. The resulting phase evolution consists of this smooth secular component, with significant phase excursions intermittently superimposed upon it.

\begin{table}
\begin{center}
  \caption{Parameters of Cubic Model for Smooth States\label{paracub}}

  \begin{tabular}{lll}
    \\
  \tableline\tableline
  \tableline\tableline
\multicolumn{3}{l}{$\phi=a_0 + a_1  (t-T_f) + a_2 (t-T_f)^2 + a_3 (t-T_f)^3$\tablenotemark{a}} \\
\multicolumn{3}{l}{$T_0 = T_f + a_0 P_f$\tablenotemark{b}} \\
\multicolumn{3}{l}{$P_{prec,0} = P_f /(1-a_1 P_f)$} \\
\multicolumn{3}{l}{${\dot P}_{prec,0} = 2 a_2 {P^2_{prec,0}}$ } \\
\multicolumn{3}{l}{${\ddot P}_{prec} = 6 a_3 {P^2_{prec,0}}$} \\
\tableline
Parameter& & Value \\
$a_0$ (cycle) &  &$0.1025 \pm 0.0040$\\ 
$a_1 $ (cycle/day) & &$(7.43 \pm 0.29) \times 10^{-5}$ \\ 
$a_2 $ (cycle/day$^2$) & &$(-1.6611 \pm 0.0031) \times 10^{-8}$ \\
$a_3 $ (cycle/day$^3$) & &$(-6.07 \pm 0.97) \times 10^{-13}$ \\
$cov(a_0, a_1)$ (cycle$^2$/day) & &   $-3.26 \times 10^{-9}$\\
$cov(a_0, a_2)$ (cycle$^2$/day$^2$)& &$-8.91 \times 10^{-13}$\\
$cov(a_0, a_3)$ (cycle$^2$/day$^3$)& &$1.68 \times 10^{-16}$\\
$cov(a_1, a_2)$ (cycle$^2$/day$^3$)& &$ 3.21 \times 10^{-16}$\\
$cov(a_1, a_3)$ (cycle$^2$/day$^4$)& &$ -1.96 \times 10^{-19}$\\
$cov(a_2, a_3)$ (cycle$^2$/day$^5$)& &$ -2.00 \times 10^{-23}$\\

$T_0$ (MJD) & &$54739.27 \pm 0.14$\\
$P_{prec,0}$ (days)& &$34.8839 \pm 0.0027$\\
${\dot P}_{prec,0}$ (day/day)& &$(-4.032 \pm 0.074) \times 10^{-5}$\\
$\ddot P_{prec}$ (day/day$^2$)& &$(-4.43 \pm 0.71) \times 10^{-9}$\\

\tableline\tableline

  \end{tabular}
  \tablenotetext{}{\parbox[t]{0.97\columnwidth}{\raggedright \bf Notes:}}
  \tablenotetext{a}{\parbox[t]{0.92\columnwidth}{\raggedright $T_f$=MJD 54726.703}}
  \tablenotetext{b}{\parbox[t]{0.92\columnwidth} {\raggedright $P_f$=34.7937 days}} 
\end{center}
\end{table}

From the onset times of the variable epochs (Epochs 2, 4, and 6) listed in Table~\ref{spve}, we found that the time interval between Epochs 2 and 4 was 3543 days, whereas that between Epochs 4 and 6 was 3241 days. The comparable intervals of 3543 and 3241 days suggest a characteristic timescale of approximately 3400 days for the recurrent phase excursions. This recurrence may indicate a long-term process that influences the precessing warped disk and produces variations in its azimuthal precessional angular velocity. However, the observed phase excursions do not persist indefinitely; instead, the disk appears to gradually recover toward its baseline long-term phase evolution, suggesting the presence of internal restoring processes. Furthermore, Epochs 4 and 6 exhibited distinct variation behaviors. As shown in Figure~\ref{phev}, the superorbital phase in Epoch 4 initially decreased smoothly and then recovered toward the long-term trend. In contrast, Epoch 6 exhibited a rapid phase ascent followed by a sharp decline that overshot below the cubic trend, before undergoing a secular relaxation back toward the long-term trajectory. A more detailed discussion of the physical mechanisms responsible for these superorbital phase variations, including the short-term fluctuations ($\sim1000$ days) and the long-term evolution ($\sim10000$ days) of the superorbital phase, is presented in Section~\ref{dis_spe}.

\subsection{Comparison Between Main High State and Short High State}\label{compms} 

Although the time resolution was lower, more precise modulation profiles could be obtained; thus, numerous SHS profiles were significantly detected from the four-supercycle light curve segments obtained with high-sensitivity instruments, including RXTE ASM, Swift BAT, and MAXI GSC. This allowed us to further investigate the properties of the SHS and its relation to the MHS. To accurately determine the parameters and profiles of the SHS, only data segments with an SHS peak height exceeding a ($5 \sigma$) detection threshold were analyzed in this study. Similar to the MHS, the peak position derived from the best-fit four-component sinusoidal model was defined as the phase of the SHS. Finally, we compared each MHS with its immediately following SHS.

\begin{figure}[htbp]
  \centering
  \includegraphics[width=0.45\textwidth]{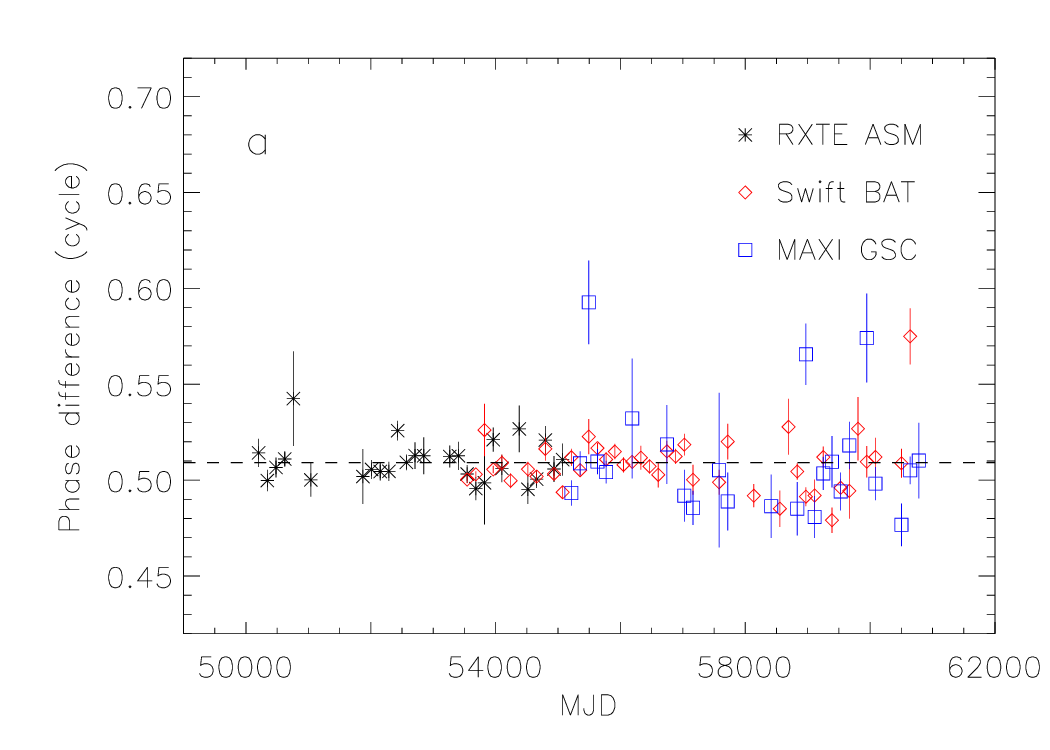}
  \includegraphics[width=0.45\textwidth]{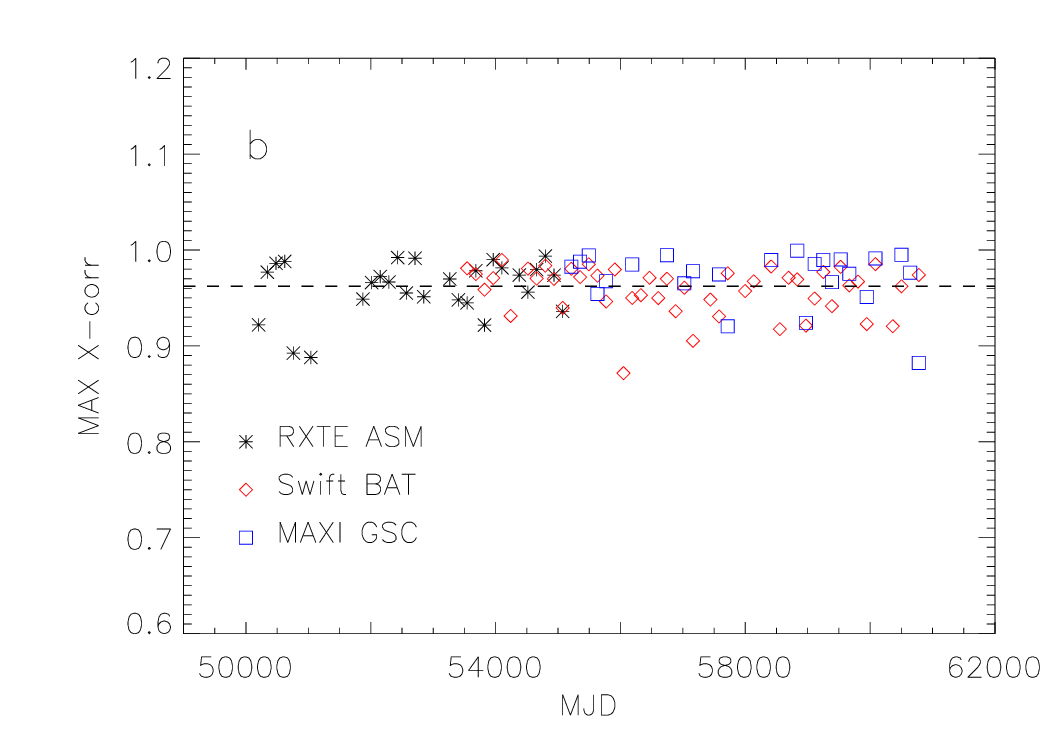}
   \includegraphics[width=0.45\textwidth]{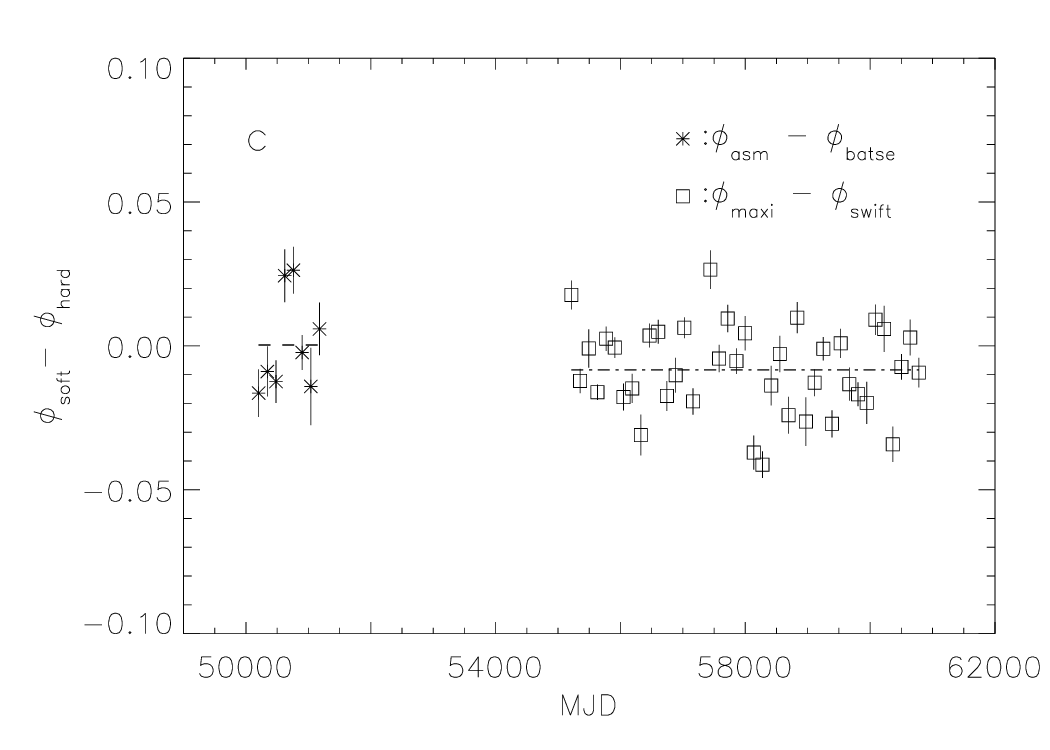}

\caption{(a) The superorbital phase differences between the SHS and MHS. No significant temporal trend is observed, with a mean phase difference of $0.509 \pm 0.017$ cycles, indicated by the dashed horizontal line. (b) The maximum normalized cross-correlation coefficients between the MHS and SHS profiles. No significant temporal trend is observed, with a mean value of $0.962 \pm 0.027$, indicated by the dashed horizontal line. (c) The superorbital phase differences obtained from overlapping soft- and hard-X-ray observations. The dashed horizontal line in the left group indicates the mean phase difference of $(-3 \pm 170) \times 10^{-4}$ cycles between RXTE ASM and CGRO BATSE, while that the dash-dotted line in the right group indicates the mean phase difference of $(-8 \pm 15) \times 10^{-3}$ cycles between MAXI GSC and Swift BAT. \label{ms_comp}}
\end{figure}

The phase differences, defined as the SHS phase minus the MHS phase ($\phi_{\rm SHS}-\phi_{\rm MHS}$), are plotted against time in the top panel of Figure~\ref{ms_comp}. The distribution is closely clustered around 0.5, with a mean value of $0.509 \pm 0.017$ and no significant trend or energy band dependence. Notably, the fluctuations increase after MJD 56500 for the Swift BAT data. An inspection of the original Swift BAT light curve reveals that this increased fluctuation is likely due to a loss of sensitivity in the instrument, which led to larger error bars after MJD $\sim56500$. We further verified the similarity of the modulation shapes between the MHS and SHS. To do this, we extracted $\pm0.2$ cycles around their respective peaks from the optimally fitted modulation profiles and performed a cross-correlation analysis between the two. The maximum normalized cross-correlation value (ranging from $-1$ to $1$) was used as a proxy for the degree of similarity between these profiles. The middle panel of Figure~\ref{ms_comp} shows the maximum normalized cross-correlation values over time. Similar to the phase differences, these values are tightly distributed around 0.95 with minor fluctuations and a mean of $0.962 \pm 0.027$, indicating that the MHS and SHS profiles are highly similar. Likewise, no significant trend or energy band dependence is evident. From the phase differences and the high degree of similarity between the MHS and SHS, we inferred the long-term geometric stability of the warped accretion disk in Her X-1, which is discussed further in Section~\ref{dis_stdisk}.

\begin{figure}
  \epsscale{1.25}
\plotone{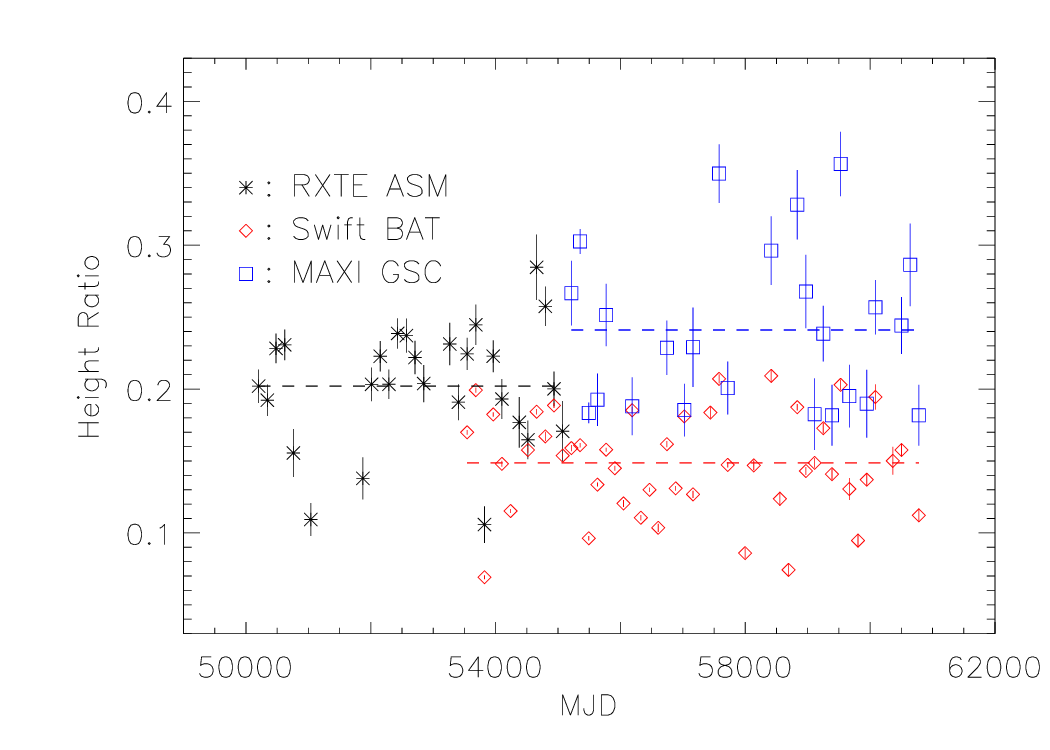}
\caption{The SHS-to-MHS height ratios. The dashed horizontal lines indicate the mean values obtained from the RXTE ASM data, $(20.2 \pm 4.2)\%$ (black); Swift BAT, $(14.9 \pm 3.5)\%$ (red); and MAXI GSC, $(24.1 \pm 5.6)\%$ (blue).\label{ms_h_ratio}}
\end{figure}

Furthermore, we evaluated the SHS-to-MHS height ratio. The ratios are $(20.2 \pm 4.2)\%$, $(14.9 \pm 3.5)\%$, and $(24.1 \pm 5.6)\%$ for the RXTE ASM, Swift BAT, and MAXI GSC observations, respectively (see  Figure~\ref{ms_h_ratio}). These results suggest a tendency toward a larger height ratio at softer X-ray energies. This tendency is qualitatively consistent with the concentric flux-contrast geometry proposed in this work. Finally, we investigated whether a significant phase difference exists between the soft and hard X-ray bands. We extracted the phases from overlapping observations of CGRO BATSE / RXTE ASM and Swift BAT / MAXI GSC. The resulting phase differences, shown in the bottom panel of Figure~\ref{ms_comp}, are roughly randomly distributed around zero. The mean phase difference is $(-3 \pm 170) \times 10^{-4}$ cycles for the CGRO BATSE / RXTE ASM dataset and $(-8 \pm 15) \times 10^{-3}$ cycles for the Swift BAT / MAXI GSC dataset. Because the measured phase differences are consistent with zero, we conclude that no significant phase difference is detected between the soft and hard X-ray bands, with a $2\sigma$ upper limit of 0.023 cycles.

\section{Discussion} \label{dis}
\subsection{Physical Mechanisms Driving the Anomalous Low State} \label{dis_als}

\subsubsection{The Two-Component Geometric Screening Model}\label{l_rank}

We proposed a non-linear geometric screening model to provide a physical basis for defining the ALS thresholds through the detection of gaps in the  MHS height rank spectrum (see Section~\ref{ials}). In this framework, the central X-ray emission zone is conceptualized as a two-component concentric structure dominated by a severe flux contrast: a primary high-flux compact region emitting direct X-rays from the neutron star, enveloped by a residual low-flux extended region (i.e. ADC) that yields much weaker, scattered photons~\citep{abd15}. Under this formulation, the physical transition between different superorbital states is regulated by the progression of the disk occultation front across these two distinct emission regions:

\begin{itemize}     

\item \textbf{Normal State:} The baseline disk tilt or warp configuration only marginally occults the outer edge of the residual low-flux extended region. Because the surface brightness of this extended scattering zone is spatially diffuse and its flux contribution is relatively low, small stochastic fluctuations in the geometric angles (such as tilt or warp variations) yield only minor, continuous variations in the MHS height ($h_{MHS}$), forming the smoothly declining linear continuum at high ranks.

\item \textbf{The First Gap and Transition State:} As the structural perturbation drives the disk edge to reach the boundary between the low-flux outer region and the primary high-flux compact region, the geometric sensitivity increases sharply. Due to the strong brightness contrast and compactness of this direct core emission, a small increase in the disk warp or tilt angle triggers a rapid occultation. In the mathematical formulation of the rank spectrum, this  non-linear screening effect implies that the flux derivative with respect to the geometric angle, $\vert{}dh_{MHS}/d\theta\vert{}$, spikes dramatically, driving the probability density of observing such intermediate fluxes ($P(h_{MHS}) \propto \vert{}d\theta/dh_{MHS}\vert{}$) close to zero. This rapid transition manifests as the first prominent gap in the rank spectrum, marking the entrance into the "transition state" where the high-flux compact region is partially obscured and the flux variation trend shifts. Based on our empirical gap identification (Table~\ref{threshold}), this critical transition boundary defines the initial ALS candidate thresholds at 18.42 mCrab for CGRO BATSE, 8.70 mCrab for RXTE ASM, and 17.53 mCrab for Swift BAT.

\item \textbf{The Second Gap and Deep Low State:} A secondary geometric boundary arises when the primary high-flux compact region approaches complete obscuration. The occultation of the remaining direct X-ray emission from this high-flux region produces another non-linear decrease in flux, forming the second distinct gap at 7.25 mCrab, which is detected only in the softer-band RXTE ASM data. Beyond this threshold, the system enters the deep low state ($h_{MHS} \sim 0$), in which the high-flux region is fully screened across all phases, and only residual coronal scattering from the extended low-flux region remains detectable, producing a distinct, flattened trend.
\end{itemize}

Consequently, our empirical methodology of identifying thresholds by searching for gaps has a plausible physical basis. The detected gaps do not represent arbitrary divisions, but can be interpreted as corresponding to the structural boundaries where the precessing, warped disk begins or approaches completion of the geometric occultation of the primary high-flux compact region. Furthermore, this two-component framework naturally accounts for the instrument-dependent differences in the rank profiles: while softer-band detectors such as RXTE ASM can resolve both geometric boundaries, yielding two distinct gaps, harder-band instruments such as CGRO BATSE and Swift BAT, which are more strongly dominated by direct emission from the primary high-flux region, primarily reveal the pronounced flux decrease associated with the occultation of the compact region.

It is worth noting that within the current observational span of Swift BAT, our rank-searching methodology has not identified a distinct deep low state (or a secondary gap) among the observed ALS episodes. However, this absence does not definitively rule out the physical existence of such a deep low regime in the hard X-ray band. Rather, it likely reflects the limited evolutionary coverage of the observed events. Given the high sensitivity of Swift BAT, a secondary gap may still become detectable if a sufficiently deep and prolonged low state occurs.

In this context, the ongoing ALS 2025 episode\footnote{Although observations after MJD 61068 (2026 January 28) are not included in the present analysis, subsequent public MAXI GSC monitoring indicates that Her X-1 has remained in the low state, with no clear evidence of recovery as of 2026 September. These post-cutoff data are noted here only for completeness and are not used in any analysis presented in this work.}, which is classified as an active candidate with a duration already exceeding four supercycles (Table~\ref{als_tb})---presents an intriguing possibility. If ALS 2025 continues to evolve toward a deeper low state, the primary high-flux compact region may eventually become completely occulted. Whether such an evolution will occur cannot yet be determined within the observational baseline adopted in this study. Future hard-X-ray monitoring may reveal a secondary flux decrease (the second gap) associated with the complete occultation of the direct X-ray-emitting region, thereby providing a direct observational test of our two-component geometric model in the hard-X-ray regime.

\subsubsection{Viscous Timescale and Evolution of ALS Durations}\label{asl_dur}

The geometric configuration of ALSs in Her X-1 remains a subject of ongoing debate. Early models by~\citet{cob00} and~\citet{cla03} suggested that these superorbital flux drops are driven by temporary changes in disk inclination (tilt). Specifically, \citet{sha98} and \citet{cob00} suggested that a lower accretion rate reduces the disk's tilt angle, causing its outer edges to block the X-ray source. Conversely, \citet{lea10a} proposed instead that an enhanced disk warp is responsible, arguing that an increased tilt is less geometrically or dynamically effective than an increased warp. In this work, ALS 1991, ALS 2015, and ALS 2025 are classified as ``ALS candidates'' because our analysis relies primarily on MHS profiles and cannot independently verify the concurrent behaviors required by the historical definition of~\citet{vrt96}.

Across our 35-year baseline (Table~\ref{als_tb}), these ALS episodes display a wide range of durations, from the two-supercycle ALS 1993 to longer episodes lasting approximately five supercycles or more (e.g., ALS 1999 and 2003). Crucially, our analysis reveals no significant correlation between the occurrence of these ALS episodes and the long-term superorbital phase variations. This lack of correlation suggests that ALS occurrence is not simply determined by the external mass-transfer fluctuations inferred from the long-term phase evolution (see Section~\ref{spe}). Instead, considering that the disk is likely inherently twisted and warped even in its steady precessional state~\citep{pri96,ogi01}, we consider a plausible framework in which ALS episodes may arise from intrinsic alterations of this pre-existing warp configuration. Under this tentative interpretation, the diverse range of observed ALS durations may reflect the extent to which such geometric perturbations can evolve and propagate through the disk relative to its characteristic viscous timescale.

According to Eq. (5.69) in~\citet{fra02}, the viscous timescale can be expressed as:

\begin{eqnarray}\label{tvisc}
 {t_{\rm visc} \sim 3\times 10^5 \alpha^{-4/5}}
\left({{\dot M} \over {10^{16}\ {\rm g\ s}^{-1}}}\right)^{-3/10}\\
\left({{m_1} \over M_{\sun}}\right)^{1/4}
\nonumber \left({R \over {10^{10}\ {\rm cm}}}\right)^{5/4}\ {\rm s}.
\end{eqnarray}

\noindent Here, $\alpha=0.3$~\citep{kin13} is the viscosity parameter, $\dot M$ represents the mass transfer rate from the companion star, $m_1=1.5M_{\sun}$ is the neutron-star mass, and $R=2.1\times10^{11}$ cm~\citep{lea25} is the disk radius. For the mass transfer rate, we also account for the ionized accretion-disk wind in Her X-1 discovered by~\citet{kos20}, whose outflow rate is approximately 70\% of the mass accretion rate ($\dot M_{\rm acc}$). Therefore, the total mass transfer rate supplied to the disk can be estimated as $\dot M \approx \dot M_{\rm acc}+\dot M_{\rm wind}=1.7\dot M_{\rm acc}$. Adopting $\dot M_{\rm acc}=1.6\times10^{17}$ g s$^{-1}$~\citep{kos26}, we obtain $\dot M=2.72\times10^{17}$ g s$^{-1}$. Substituting these parameters into Eq.~\ref{tvisc} yields a characteristic viscous timescale of $t_{\rm visc}\approx 168.0$ days.

This estimated viscous timescale provides a physically motivated timescale against which the observed ALS durations can be compared. The value $t_{\rm visc}\approx 168.0$ days corresponds to approximately 4.8 superorbital cycles, comparable to the duration of the intermediate-duration ALS episodes. We therefore tentatively interpret the observed ALS durations in terms of three phenomenological regimes according to their relation to $t_{\rm visc}$. For brief one-cycle events and the two-cycle ALS 1993 episode ($t_{\rm ALS}<t_{\rm visc}$), an intrinsic geometric perturbation may dissipate or weaken before viscous communication extends throughout the disk. Nevertheless, if such a perturbation occurs within a favorable pre-existing warp geometry, even a relatively small and short-lived alteration may be sufficient to move the line of sight into the occulted region, producing a rapid flux decrease that reaches only the transition state without requiring a global reconfiguration of the disk.

For ALS episodes lasting approximately five supercycles ($t_{\rm ALS}\sim t_{\rm visc}$), the perturbation persists for approximately one viscous timescale and may therefore have sufficient time to affect a substantial fraction of the disk, producing a more persistent alteration of the warp configuration. In contrast, the historical ALS 1999 (16 cycles) and ALS 2003 (8 cycles) episodes ($t_{\rm ALS}>t_{\rm visc}$) persisted for multiple viscous timescales, potentially allowing the perturbation to propagate more extensively through the disk and produce a more global reconfiguration of its warped structure. Such long-lived geometric changes may account for the prolonged transition and deep low states, during which the central X-ray source becomes strongly or completely occulted over the superorbital cycle. In this picture, the estimated 168-day viscous timescale does not represent a strict threshold for ALS evolution, but rather a characteristic communication timescale that provides a natural framework for interpreting the wide range of observed ALS durations.

\subsection{Superorbital Phase Evolution} \label{dis_spe}

\subsubsection{Secular Trends and Recurrent Variable Epochs} \label{ph_var}

The 35-day superorbital modulation of Her X-1 in the X-ray band is widely attributed to the periodic occultation of X-rays emitted from the central neutron star and inner accretion disk by a precessing, warped, and tilted disk, driven by tidal interactions with the companion star, HZ Her. Observed variations in the superorbital phase allow us to probe fluctuations in the disk precession rate, providing a powerful diagnostic for investigating accretion disk dynamics. In this section, we explore whether these variations in the superorbital phase can be interpreted in terms of changes in the accretion disk mass.

 To evaluate these dynamics analytically, we employed the classic formula derived for a rigidly precessing, tilted disk as a first-order approximation. Although the accretion disk in Her X-1 is structurally warped, strong internal hydrodynamic coupling and pressure forces are expected to force the disk to precess as a coherent, unified body~\citep{lub00}. Under this regime of rigid co-precession, the complex, radius-dependent warping profile can be effectively averaged out, allowing the global tidal precession rate to be reliably approximated by an equivalent tilted-disk framework. Following Eq. 21 in \citet{lar96} or Eq. 1 in \citet{lar98}, the global disk precession angular velocity can be expressed as:

\begin{equation}\label{preangv} 
\Omega_{prec} =- {3 \over 4}{{Gm_2} \over a^3} {{\int_{R_{in}}^R \Sigma(r) r^3 dr} \over{\int_{R_{in}}^R \Sigma(r) \Omega(r) r^3 dr}}  \cos(\delta)
\end{equation}

\noindent where $m_{2}$ is the mass of the companion star, $a$ is the binary separation, and $R_{\mathrm{in}}$ and $R$ represent the inner and outer radii of the accretion disk, respectively. Additionally, $\Sigma(r)$ denotes the surface density profile, $\Omega(r)= \sqrt{Gm_1/r^3}$ is the Keplerian angular velocity (where $m_1$ is mass of neutron star in Her X-1), and $\delta$ is the disk inclination angle relative to the binary orbital plane. 

It should be noted that variations in the disk mass are not the only possible mechanism capable of modifying the global precession rate. As indicated by Equation~(\ref{preangv}), changes in the radial surface-density distribution $\Sigma(r)$ or in the disk inclination angle $\delta$ can also alter $\Omega_{\rm prec}$, even without a substantial change in the total disk mass. Furthermore, because the actual accretion disk in Her X-1 is warped, temporal variations in the radial warp or twist profile may modify the effective torque and angular-momentum distribution of the disk and thereby affect its global precession rate. These effects are not explicitly included in the rigidly precessing tilted-disk approximation adopted here. Therefore, the disk-mass variations inferred below should be regarded as a plausible first-order interpretation of the observed phase evolution rather than a unique physical solution.

We first assume that the inclination angle $\delta$ is a constant and that the accretion disk can be described by a gas-pressure-dominated $\alpha$-disk with Kramers opacity. According to Eq.~(5.49) in~\citet{fra02}, $\Sigma (r) \propto r^{-3/4}$ and can be parameterized as:

\begin{equation}\label{surden} 
  \Sigma (r)= \Sigma_0 {({r \over R})^{-3/4}}
\end{equation}

\noindent where $\Sigma_0$ is the surface density at the outer radius $r=R$.   Assuming that $R_{in} \ll R$, Eq.~\ref{preangv} yields the relation $\Omega_{prec} \propto R^{3/2}$. Under this profile, the total mass of the accretion disk is given by:

\begin{equation}\label{mvsr} 
  M_{disk}=\int_0^R 2\pi r \Sigma(r) dr = {{8 \pi} \over 5} \Sigma_0 R^2
\end{equation}

\noindent If we assume that variations in the disk mass are primarily manifested through variations in the outer disk radius \(R\) while \(\Sigma _{0}\) remains relatively steady,  we can infer that $\Omega_{prec} \propto M_{disk}^{3/4}$ , or equivalently, the precession period scales as $P_{prec} \propto M_{disk}^{-3/4}$.  Differentiating this relation yields the fractional change rates:

\begin{equation}\label{ratiochange} \frac{\dot{M}_{\mathrm{disk}}}{M_{\mathrm{disk}}} = -\frac{4}{3} \frac{\dot{P}_{\mathrm{prec}}}{P_{\mathrm{prec}}} \end{equation}

Based on the cubic-trend parameters listed in Table~\ref{paracub}, the precession period, $P_{\mathrm{prec}}$, and its first-order derivative, $\dot{P}_{\mathrm{prec}}$, both exhibit monotonic evolution over the interval from MJD 49925 to 60780. Specifically, $P_{\mathrm{prec}}$ decreases from $35.0267 \pm 0.0094$ days to $34.560 \pm 0.014$ days, corresponding to a fractional decrease of  $\sim 1.33\%$ over $\sim30$ years. This change is larger than the $0.55\%$ variation in the superorbital period of LMC X-4 reported over a 33-year baseline~\citep{chou26}. Over the same interval, $\dot{P}_{\mathrm{prec}}$ becomes more negative, changing from $(-1.90 \pm 0.35) \times 10^{-5}$~d~d$^{-1}$ to $(-6.71 \pm 0.44) \times 10^{-5}$~d~d$^{-1}$. By substituting the time-dependent evolution $\dot{P}_{\mathrm{prec}}(t) = \dot{P}_{\mathrm{prec},0} + \ddot{P}_{\mathrm{prec}}(t-T_0)$ into Eq.~\ref{ratiochange} and integrating over this 29.7-year baseline, the adopted first-order disk model implies a cumulative increase of $(1.790 \pm 0.036)\%$ in the total accretion-disk mass. Furthermore, given the scaling relation $M_{\mathrm{disk}} \propto R^2$ from Eq.~\ref{mvsr}, the corresponding outer-disk radius is inferred to have expanded by $(0.895 \pm 0.018)\%$ over the same interval.

To explain how the system alters its global angular momentum budget and drives the observed precession drift, we explored the potential physical origins of the $\sim 1.8\%$ cumulative mass enhancement. In a steady-state framework, the mass budget of the precessing disk is governed by

\begin{equation}\label{diskmasschange} 
 {\dot M}_{disk} = {\dot M}_{trans} - {\dot M}_{acc} - {\dot M}_{wind}
\end{equation}

\noindent where ${\dot M}_{trans}$ is the mass transfer rate from the companion, ${\dot M}_{acc}$ is the accretion rate, and ${\dot M}_{wind}$ is the mass outflow rate of the disk wind, which is $\sim70\%$ of ${\dot M}_{acc}$~\citep{kos20}. Under this mass-budget framework, three possible scenarios could account for the required $1.8\%$ secular accumulation. First, a marginal enhancement of $\sim 1$--$2\%$ in the companion mass transfer rate ${\dot M}_{\mathrm{trans}}$ over the 29.7-year baseline, possibly associated with long-term thermodynamic adjustments or atmospheric perturbations of HZ Her, could lead to mass accumulation in the disk. Second, a small decrease in the core accretion rate ${\dot M}_{\mathrm{acc}}$ due to a secular change in the internal disk viscosity could also reduce the outflow rate, effectively retaining a larger fraction of the transferred material within the disk. Third, the intrinsic efficiency of the disk wind may have decreased slightly from a higher historical level to the currently observed $70\%$ level. Given the small magnitude of this $1.8\%$ alteration, these three scenarios remain observationally degenerate; nevertheless, any combination of them could potentially account for the inferred $0.9\%$ expansion of the outer disk radius, providing a self-consistent interpretation of the accelerated 35-day superorbital modulation.

We suggest that the same mechanism may explain the phase variations observed during the variable epochs (Epochs 2, 4, and 6, as defined in Section~\ref{spe} and listed in Table~\ref{spve}). We attempted to evaluate the changes in accretion disk mass during these periods. However, because the phase variation for Epoch 2 was incomplete due to ALS 1999, Epochs 4 and 6 were adopted for the evaluation instead. To isolate the disk-mass change from long-term evolution, we analyzed the residual phases after subtracting the cubic trend (see Figure~\ref{ph_res_fit}).  Unlike the long-term evolution, we lack a robust model for describing the phase variations during these variable epochs; therefore, the analysis below represents only a first-order approximation and provides an order-of-magnitude estimate. As listed in Table~\ref{spv46} and shown in Figure~\ref{ph_res_fit}, Epoch 4 was divided into two stages based on its phase variation trend, termed the "Decline'' and "Recovery'' stages. Meanwhile, Epoch 6 was divided into three stages: "Fast Ascent'', "Severe Decline'', and "Secular Relaxation''. Integrating Eq.~\ref{ratiochange}, we obtained:

\begin{figure}
  \epsscale{1.25}
\plotone{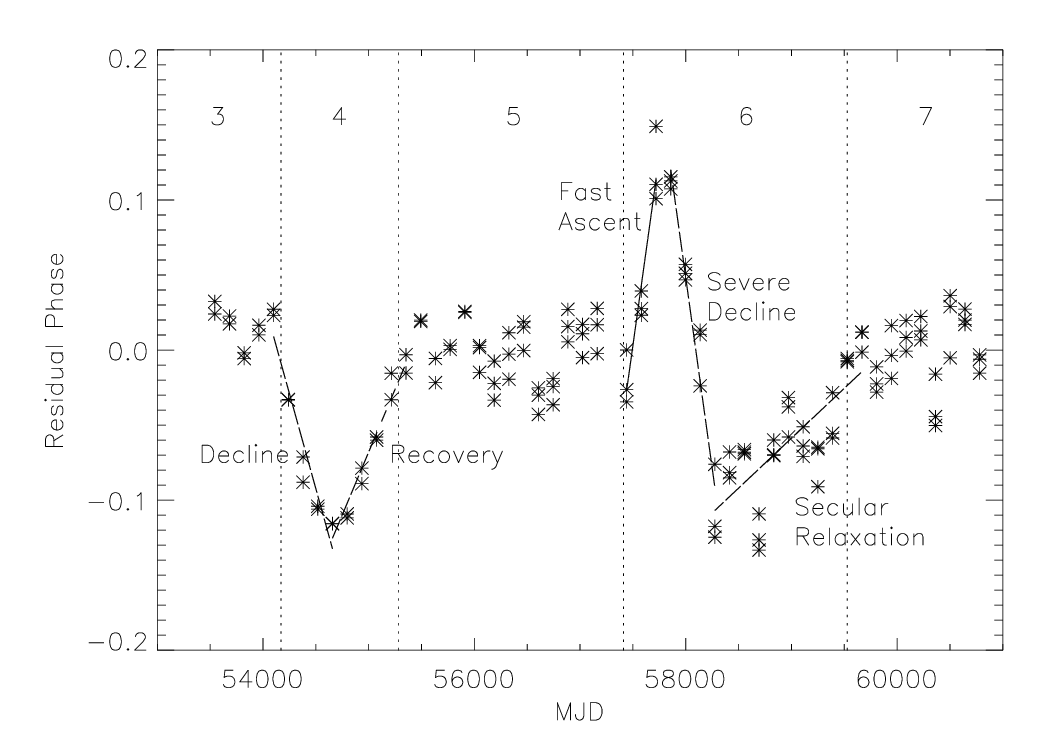}
\caption{The superorbital phase residuals (see Figure~\ref{phev}) from the end of Epoch 3 to the beginning of Epoch 7. Variable Epoch 4 is divided into two stages, whereas variable Epoch 6 is divided into three stages according to their phase-variation patterns. The slope obtained by fitting a linear function to each stage is used to estimate the corresponding fractional change in the accretion-disk mass within the adopted first-order model framework (see Section~\ref{ph_var}).
\label{ph_res_fit}}
\end{figure}

\begin{table*}[ht]
\centering
  \caption{Superorbital Phase Variations in Epochs 4 and 6\label{spv46}}

  \begin{tabular}{lccccc}
    \\
    \tableline\tableline
     \tableline\tableline
Epoch/Stage & Start time\tablenotemark{a} & Stop time\tablenotemark{a} &$\Delta \phi/\Delta t$\tablenotemark{b} & ${\Delta M_{disk}}/M_{disk}$\tablenotemark{c} & ${\Delta R}/R$\\
             & (MJD)     & (MJD)        & (cycle/day)        & (\%) & (\%)            \\ 
\tableline
  \tableline
  Epoch 4  & & & &\\
   \tableline
   Stage 1: Decline & 54100 & 54657   & $(-2.53 \pm 0.26) \times 10^{-4}$ & $+1.18 \pm 0.12$ & $+0.589 \pm 0.060$ \\
   Stage 2: Recovery & 54657 & 55353 &$(1.67 \pm 0.11) \times 10^{-4}$ & $-0.776 \pm 0.051$& $-0.388 \pm 0.026$\\
   \tableline
     \tableline

  Epoch 6  & & & \\
   \tableline
   Stage 1: Fast Ascent & 57411& 57858  &$(5.05 \pm 0.60)\times 10^{-4}$ & $-2.34 \pm 0.28$ & $-1.17 \pm 0.14$\\
   Stage 2: Severe Decline & 57858& 58276  &$(-5.07 \pm 0.39) \times 10^{-4}$ & $+2.35 \pm 0.18$& $+1.175 \pm 0.091$\\
   Stage 3: Secular Relaxation & 58276 & 59567 & $(6.65 \pm 0.98) \times 10^{-5}$& $-0.308 \pm 0.046$ & $-0.154 \pm 0.023$\\
\tableline\tableline
\tableline\tableline
  \end{tabular}
  \tablenotetext{}{\parbox[t]{1.94\columnwidth}{\raggedright \bf Notes:}}
  \tablenotetext{a}{\parbox[t]{1.92\columnwidth} {\raggedright The time intervals listed here correspond to the phase-residual data used for the linear fits and may therefore extend slightly beyond the nominal epoch boundaries in Table~\ref{spve}. Each phase measurement represents a four-supercycle segment and is assigned to its midpoint. }}
   \tablenotetext{b}{\parbox[t]{1.92\columnwidth} {\raggedright Evaluated by fitting the phase residuals (Figure~\ref{ph_res_fit}) in the corresponding time intervals with a linear function.}}
  \tablenotetext{c}{\parbox[t]{1.92\columnwidth} {\raggedright Assuming the disk surface density $\Sigma(r) \propto r^{-3/4} $}}

\end{table*}

\begin{equation}\label{ratiochange1} \frac{\Delta {M}_{\mathrm{disk}}}{M_{\mathrm{disk}}} = -\frac{4}{3} \frac{\Delta{P}_{\mathrm{prec}}}{P_{\mathrm{prec}}}=\frac{4}{3} \frac{\Delta{\nu}_{\mathrm{prec}}}{\nu_{\mathrm{prec}}} \end{equation}

\noindent under the assumption that ${\Delta{P}_{\mathrm{prec}}}/{P_{\mathrm{prec}}} \ll 1$. For each stage, we fitted a linear function to obtain the mean phase change rate ($\Delta \phi /\Delta t$), yielding
\begin{equation}\label{phisloop} 
{{\Delta \phi} \over {\Delta t}}= {-\Delta \nu_{prec}}=-\frac{3}{4} \frac{\Delta {M}_{\mathrm{disk}}}{{M}_{\mathrm{disk}}} \nu_{prec}
\end{equation}

\noindent or

\begin{equation}\label{phisloop1} 
 \frac{\Delta {M}_{\mathrm{disk}}}{{M}_{\mathrm{disk}}} =-\frac{4}{3} \frac{\Delta \phi}{\Delta t}  P_{prec} 
\end{equation}

\noindent We fitted the phase residuals with a linear function to obtain the $\Delta \phi / \Delta t$ values shown in Figure~\ref{ph_res_fit}. The evaluated ${\Delta {M}{\mathrm{disk}}}/{{M}{\mathrm{disk}}}$ and $\Delta R/R$ values are listed in Table~\ref{spv46}. In principle, the total ${\Delta {M}_{\mathrm{disk}}}/{{M}_{\mathrm{disk}}}$ over an entire epoch should be zero, as the phases eventually return to their normal state (i.e., zero residual phase) if a sufficiently robust model is available to describe the phase evolution in Epochs 4 and 6. However, because our estimation is only a first-order approximation, the resulting total ${\Delta {M}_{\mathrm{disk}}}/{{M}_{\mathrm{disk}}}$ does not equal zero for each epoch. Nevertheless, these estimates should be reliable to within an order of magnitude, suggesting that a small change in disk mass ($\vert{\Delta {M}_{\mathrm{disk}}}/{{M}_{\mathrm{disk}}}\vert \lesssim 2.4\%$) can induce significant phase modulation during the variable epochs, including Epochs 4 and 6 and potentially Epoch 2, on timescales of 1000--2000 days.

\subsubsection{Fluid Disk Micro-Perturbation Model and Viscous Relaxation}\label{mag_per}

We propose a scenario in which the disk-mass changes during the variable epochs may result from variations in the mass transfer rate from the companion through the $L_1$ region, driven by the stellar magnetic cycle of HZ Her. As the results presented in Section~\ref{spe} indicate, the onset times of these variable epochs appear to recur quasi-periodically on a characteristic timescale of $\sim$3400~days ($\sim$9.3~years), close to the $\sim$10-year periodicity proposed by~\citet{sta09}. Variations in the magnetic pressure around the $L_1$ point can directly affect the atmospheric scale height. For a persistent X-ray source such as Her X-1, the companion star HZ Her must continuously undergo substantial Roche-lobe overflow. In such a state, the stellar photospheric radius extends beyond the critical geometric boundary ($R_{*} > R_2$), such that the $L_1$ point is effectively embedded within the stellar atmosphere. Following Eq. (4.19) of~\citet{fra02}, the relative change in the mass transfer rate can then be expressed, to first order, as

\begin{equation}\label{rlof} 
\frac{{d\dot M}_{2}}{{\dot M}_{2}}  \approx -\left(\frac{R_{*}-R_{2}}{H_{*}}\right)\frac{dH_{*}}{H_{*}},
\end{equation}

\noindent where $H_{*}$ is the atmospheric scale height. To explicitly account for the role of magnetic activity, the total pressure must include both gas pressure and magnetic pressure. Consequently, the magnetically modified atmospheric scale height can be expressed as:

\begin{equation}\label{hmag}
H_{*} = H_0\left(1+\frac{B^2}{8\pi P_{\rm gas}}\right),
\end{equation}

\noindent where $H_0$ is the unperturbed gas scale height, $B$ is the local magnetic field strength, and $P_{\rm gas}$ is the gas pressure near the L1 region. Crucially, because $R_{*} > R_2$, the factor $(R_{*}-R_2)/H_{*}$ in Eq.~\ref{rlof} is positive. It follows from Eq.~\ref{hmag} that a localized weakening of the surface magnetic field ($dB<0$) reduces the magnetic pressure and hence decreases the atmospheric scale height ($dH_{*}<0$). According to Eq.~\ref{rlof}, the negative sign then implies that it leads to $d\dot{M}_2>0$, corresponding to an enhancement of the mass-transfer rate through the $ L_{1}$ region.

To compare this hypothesis with our observations, we first examined the phase variation during Epoch 4. As shown in Figure~\ref{ph_res_fit}, the phase residuals in Epoch 4 exhibited a distinct two-stage behavior. From MJD 54100 to 54657, the residuals decreased at a rate of $-2.53 \times 10^{-4}$~cycles~day$^{-1}$ (Table~\ref{spv46}). This negative phase derivative corresponds, within our model, to an accumulation of accretion disk mass. Under the magnetic-cycle scenario proposed above, this accumulation may have resulted from a sustained enhancement of mass transfer associated with a weakening of the magnetic field ($dB < 0$), which would decrease $H_{*}$ and increase ${\dot M}_{2}$ above its baseline level. This mass-injection phase lasted for 557~days, corresponding to approximately $3.3\times t_{\mathrm{visc}}$ based on our estimated disk viscous timescale of $t_{\mathrm{visc}}\sim168.0$ days (see Eq.~\ref{tvisc} and Section~\ref{dis_als}). Subsequently, from MJD 54657 to 55353, the trend reversed, and the residuals increased at a rate of $1.67 \times 10^{-4}$~cycles~day$^{-1}$, gradually returning to the baseline level (where the residual equals zero) over a duration of 696~days. We interpret this recovery as the possible viscous response of the disk after the magnetic perturbation subsided and ${\dot M}_{2}$ returned to its baseline level, allowing the excess disk mass to be gradually redistributed and drained. This recovery duration corresponds to approximately $4.1\times t_{\mathrm{visc}}$. Such a duration is consistent with relaxation over several viscous timescales and therefore provides qualitative support for a viscous origin of the observed recovery toward the unperturbed baseline state.

A more dynamic manifestation of this scenario may be present in Epoch 6, which exhibited a multi-stage phase variation followed by a prolonged relaxation phase. As listed in Table~\ref{spv46}, Epoch 6 was divided into three distinct stages. Stage 1 (Fast Ascent) extended from MJD 57411 to 57858 with a positive slope of $5.05 \times 10^{-4}$~cycles~day$^{-1}$, corresponding, within our model, to a rapid depletion of disk mass. Under the magnetic-cycle scenario proposed above, this behavior may be associated with a localized enhancement of the stellar magnetic field ($dB > 0$), which would increase the scale height ($dH_{*} > 0$ via Eq.~\ref{hmag}), reduce the gas density at the $L_{1}$ point, and thereby decrease the mass transfer rate to the disk for approximately 447~days ($\sim 2.7 \times t_{\mathrm{visc}}$). Immediately following this, Stage 2 (Severe Decline) extended from MJD 57858 to 58276 with a nearly symmetric negative slope of $-5.07 \times 10^{-4}$~cycles~day$^{-1}$ over 418~days ($\sim 2.4 \times t_{\mathrm{visc}}$). Within our proposed scenario, this rapid reversal may correspond to a weakening of the magnetic field, which would decrease the atmospheric scale height and enhance the mass transfer rate, resulting in a rapid accumulation of disk mass. Finally, Stage 3 (Secular Relaxation) extended from MJD 58276 to 59567 (1291~days) and was characterized by a much shallower positive slope of $6.65 \times 10^{-5}$~cycles~day$^{-1}$. If the external perturbation had subsided by the onset of Stage 3, the subsequent evolution could represent the viscous redistribution and drainage of the excess disk mass accumulated during Stage 2. The duration of this secular relaxation phase corresponds to $1291/168.0 \approx 7.7 \times t_{\mathrm{visc}}$. Together with the $\sim4.1\times t_{\mathrm{visc}}$ recovery timescale found for Epoch 4, this result suggests that the recovery phases in both epochs occurred over several characteristic viscous timescales. This correspondence is consistent with, although does not uniquely establish, a scenario in which variations in stellar magnetic activity modulate the mass transfer rate and subsequently drive a viscous response in the accretion disk.

Extrapolating the $\sim3400$-day characteristic timescale of the recurrent phase variations suggests that the subsequent variable epoch may have begun around MJD $\sim60811$ (May 16, 2025). However, Her X-1 entered ALS 2025 on MJD 60920, only $\sim109$ days, or approximately three superorbital cycles, after this predicted onset. This interval is shorter than the four-supercycle data segment required for an independent phase measurement in our analysis. Furthermore, the suppression of the superorbital modulation profile during ALS 2025 prevented precise phase determination thereafter. Consequently, the onset of this potential variable epoch cannot be definitively confirmed with the current dataset.

In contrast, during the smooth epochs (e.g., Epoch 3), the phase residuals remain remarkably stable near zero without exhibiting distinct multi-stage modulations. It is worth noting that even during such a smooth epoch, the accretion disk mass still undergoes a gradual, long-term variation. Under our proposed framework, the lack of short-term residual fluctuations is consistent with a relatively quiescent phase of HZ Her's stellar magnetic cycle, during which the surface magnetic field remains relatively stable ($dB \approx 0$). Consequently, the atmospheric scale height and mass transfer rate would follow the steady, long-term baseline evolution of the system without additional modulation associated with magnetic perturbations, thereby allowing the phase residuals to remain close to zero.

Regarding variations in the accretion disk radius, it is well known that the maximum extent of an accretion disk in a close binary system is limited by the tidal truncation radius, $R_{\mathrm{tid}}$, where tidal torques from the companion star effectively remove angular momentum from the disk edge~\citep[e.g.,][]{pap77}. However, $R_{\mathrm{tid}}$ does not act as a rigid structural boundary. Instead, the actual disk radius $R$ is determined by a dynamic balance between the outward viscous transport of angular momentum and tidal torques near the disk edge. Our analysis reveals that a small variation in the outer radius, specifically $\vert\Delta R / R\vert \lesssim 1.2\%$ (see Table~\ref{spv46}), is sufficient to account for the observed superorbital phase fluctuations. Such a small fractional variation can be accommodated as a perturbation of the outer disk radius without requiring the disk to extend beyond its tidal truncation limit. Variations in the viscous evolution of the outer disk or in the mass transfer rate from HZ Her could produce such low-amplitude changes in $R$, providing a plausible physical explanation for the inferred radius variations associated with the superorbital phase excursions in Her X-1.

In summary, our analytical approach yields two main insights into the superorbital dynamics of Her X-1. First, by adopting the rigidly precessing tilted-disk model as a first-order approximation, we demonstrate that a small fractional variation in the accretion disk mass (on the order of a few percent) is sufficient to account for the pronounced superorbital phase shifts observed in the data. Within the assumptions of this approximation, the corresponding changes in the outer disk radius are also small ($\lesssim 1.2\%$) and can be accommodated within the tidal truncation constraint. Second, our ability to infer such subtle, low-amplitude structural variations in the disk underscores the diagnostic value of monitoring superorbital phase evolution. Because the observed phase drift represents a cumulative, time-integrated effect of variations in the disk precession rate, it provides enhanced sensitivity to small dynamical changes in the disk that might otherwise be difficult to detect directly.

\subsection{Disk Geometric Properties: Global Structural Stability and State Height Ratio} \label{dis_stdisk}

Based on the multi-satellite timing and profile analyses presented in Section~\ref{compms}, we discuss the geometric symmetry and structural stability of the precessing accretion disk in Her X-1. Our empirical measurements provide observational constraints indicating a highly symmetric and stable global disk configuration over the 35-year baseline.

First, the superorbital phase difference between the MHS and the SHS is tightly distributed around 0.5 cycles, yielding a mean value of $0.509 \pm 0.017$ cycles. This near-antiphase alignment is consistent with an approximately symmetric geometry of the two X-ray-visible intervals produced by the precessing warped disk, suggesting a high degree of azimuthal symmetry in its global structure. Second, the maximum normalized cross-correlation coefficients between the MHS and SHS profiles are closely clustered around 0.95, with a mean of $0.962 \pm 0.027$. This high degree of profile similarity suggests that the global occultation geometry of the warped disk remains stable over the observational baseline, with only limited structural variations.

Third, our cross-instrument timing analysis reveals that the superorbital phase differences between overlapping soft- and hard-X-ray observations are consistent with zero. The measured phase offsets are $(-3 \pm 170) \times 10^{-4}$ cycles for the CGRO BATSE / RXTE ASM dataset and $(-8 \pm 15) \times 10^{-3}$ cycles for the Swift BAT / MAXI GSC dataset. The absence of a significant energy-dependent phase offset indicates that the soft- and hard-X-ray superorbital modulations trace essentially the same occultation geometry. This energy-independent timing behavior is consistent with a relatively well-defined global occulting structure in the warped disk, rather than one producing a measurable energy-dependent displacement of the occultation boundary.

Comparing these results with the 33-year multi-instrument monitoring of LMC X-4 presented by~\citet{chou26} reveals a clear difference in the long-term behavior of the two precessing disk systems. While LMC X-4 exhibited a significant superorbital phase shift of $0.044 \pm 0.010$ cycles during MJD 57000--60461, which was interpreted as a geometric transition from an approximately anti-symmetric to an asymmetric disk configuration, Her X-1 shows no significant energy-dependent phase offset and maintains a highly stable MHS--SHS phase relation and profile similarity over our multi-decadal baseline. This comparison suggests that the global warped-disk geometry of Her X-1 is substantially more stable over long timescales than that inferred for LMC X-4, despite the short-term phase excursions identified in Her X-1.

For the SHS-to-MHS height ratio, during the MHS peak, the precessing disk warp provides a relatively unobscured line of sight to the central X-ray source, allowing the direct emission from the compact region to contribute strongly, together with emission from more extended scattering and reflection components. Consequently, the denominator ($h_{\text{MHS}}$) remains relatively high across the energy bands. In contrast, during the SHS, the viewing geometry of the precessing warped disk partially occults the compact central X-ray source, reducing the contribution from the direct emission more strongly than that from the spatially extended scattering and reflection components.

Because harder X-ray instruments (e.g., CGRO BATSE and Swift BAT) are more strongly dominated by direct emission from the compact central region, partial occultation of this component during the SHS produces a relatively large reduction in the observed hard X-ray flux, resulting in a lower $h_{\text{SHS}}/h_{\text{MHS}}$ ratio. Conversely, softer X-ray instruments (e.g., RXTE ASM and MAXI GSC) may receive a relatively larger contribution from scattered and reprocessed emission associated with the extended disk corona and the companion star. Because these spatially extended components are less strongly affected by geometric occultation than the compact direct-emission region, the soft-band $h_{\text{SHS}}$ undergoes a smaller fractional reduction relative to $h_{\text{MHS}}$. The combination of stronger suppression of the direct hard X-ray component and the relatively persistent contribution of extended soft X-ray emission therefore naturally produces a larger $h_{\text{SHS}}/h_{\text{MHS}}$ ratio toward softer energy bands.  This tendency toward a larger ratio at softer energies is consistent with, and provides additional support for, the concentric flux-contrast geometry proposed in this work.

\section{SUMMARY}
We analyzed approximately 35 years of X-ray monitoring data from CGRO BATSE, RXTE ASM, Swift BAT, MAXI GSC, and Fermi GBM to investigate the long-term superorbital behavior of Her X-1. The main results of this study are summarized as follows:

\begin{itemize}

\item We developed an objective, data-driven method to identify anomalous low states (ALSs) from the distribution of main high-state (MHS) peak heights. The method successfully recovers the previously reported ALSs in 1993, 1999, and 2003, identifies two previously unrecognized candidate events, ALS 1991 and ALS 2015, and independently recovers the recently reported ALS 2025, which was still ongoing at the end of our observational baseline. The 1991 and 2015 candidates persisted for $\sim5$ superorbital cycles and remained in the transition state, whereas the longer ALSs in 1999 and 2003 evolved from the transition state into a deep low state. This provides a systematic observational framework for identifying and characterizing different stages of ALS evolution in Her X-1.

\item The superorbital phase evolution consists of a smooth secular component with intermittent phase excursions. The smooth component is empirically described by a cubic polynomial, while the variable epochs recur on a characteristic timescale of $\sim3400$ days ($\sim9.3$ yr), based on the comparable intervals of 3543 and 3241 days. Within the adopted first-order tilted-disk model, the secular evolution implies a gradual decrease in the disk precession period from $35.0267\pm0.0094$ to $34.560\pm0.014$ days, corresponding to an inferred cumulative disk-mass increase of $(1.790\pm0.036)\%$ and an outer-radius expansion of $(0.895\pm0.018)\%$.

\item Within the same model framework, the transient phase excursions can be reproduced by relatively small disk-mass perturbations, with $|\Delta M_{\rm disk}/M_{\rm disk}|\lesssim2.4\%$. Their recovery over several viscous timescales ($t_{\rm visc}\sim168$ days) is consistent with a scenario in which the disk responds to temporary mass-transfer variations and subsequently relaxes toward its secular state. The $\sim3400$-day characteristic timescale may be related to long-term variations in the mass-transfer rate from HZ Her, possibly associated with magnetic activity, although the present observations do not establish a causal connection.

\item The MHS and SHS remain highly coherent, with a mean phase separation of $0.509\pm0.017$ cycles and a mean normalized cross-correlation coefficient of $0.962\pm0.027$. No significant energy-dependent phase difference is detected, with a $2\sigma$ upper limit of 0.023 cycles. Together, these results indicate that the warped accretion disk maintains a stable global geometry over the observational baseline, while small variations in its mass and structure are sensitively traced by the long-term superorbital phase evolution.

\end{itemize}


\begin{acknowledgments}
The authors thank Dr. Mike T.-C. Yang and Mr. J.-K. Guo for their technical assistance in resolving software issues, which was essential for the data analysis. This research has made use of data and software provided by the High Energy Astrophysics Science Archive Research Center (HEASARC), which is a service of the Astrophysics Science Division at NASA/GSFC. We express our gratitude to the Gamma-Ray Astrophysics Team of the  NSSTC for archiving the CGRO BATSE and Fermi GBM data. This study also made use of data provided by the RXTE ASM team at MIT and GSFC. We also acknowledge the use of public data from the Swift data archive, as well as the  MAXI data provided by RIKEN, JAXA and the MAXI team.
\end{acknowledgments}

\section*{Disclosure of Delegation to Generative AI}
The authors declare the use of generative AI in the research and writing process. According to the GAIDeT taxonomy (2026), the following tasks were delegated to GAI tools under full human supervision:

\noindent - Text generation

\noindent - Proofreading and editing

\noindent The GAI tool used was: ChatGPT.

\noindent Responsibility for the final manuscript lies entirely with the authors.
GAI tools are not listed as authors and do not bear responsibility for the final outcomes.

\noindent Declaration submitted by: Yi Chou

\

\vspace{5mm}
\facilities{ADS, HEASARC, CGRO (BATSE), RXTE (ASM), Swift (BAT), MAXI (GSC), Fermi (GBM)}
\software{HEASoft(v6.35.1)}


%%\software{astropy \citep{2013A&A...558A..33A,2018AJ....156..123A},  
%%          Cloudy \citep{2013RMxAA..49..137F}, 
%%          Source Extractor \citep{1996A&AS..117..393B}
%%          }



\end{CJK*}

\end{document}